\documentclass[preprint,3p]{elsarticle}

\usepackage[utf8]{inputenc}
\usepackage{natbib}
\usepackage{graphicx}
\usepackage{amssymb}
\usepackage{amsmath}
\usepackage{float}\usepackage[tight,footnotesize]{subfigure}
\usepackage{color}
\usepackage{hyperref}

\journal{Journal of \LaTeX\ Templates}
\biboptions{sort&compress}

\definecolor{red-4link}{rgb}{0.62,0.03,0.}
\definecolor{green-4link}{rgb}{0.,0.67,0.}
\definecolor{blue-4link}{rgb}{0.,0.,1.}
\definecolor{orange-4link}{rgb}{0.91,0.53,0.11}

\definecolor{purple-density}{rgb}{.37,.17,.52}
\definecolor{orange-density}{rgb}{0.73,0.28,0.16}
\definecolor{brown-density}{rgb}{0.64,0.61,0.15}
\definecolor{blue-density}{rgb}{0.31,0.84,1.}

\definecolor{blue-gl}{rgb}{0., 0.561, 1.}
\definecolor{red-gl}{rgb}{0.7333, 0., 0.0901}
\definecolor{green-gl}{rgb}{0, 0.929, 0.2039}

\usepackage{ulem}
\usepackage{xcolor}

\RequirePackage{snapshot}

\begin{document}

\begin{frontmatter}

\title{Braess paradox in a network with stochastic dynamics and fixed strategies}

\author[cologne]{Stefan Bittihn\corref{cor1}}
\ead{bittihn@thp.uni-koeln.de} 
\author[cologne]{Andreas   Schadschneider} 
\ead{as@thp.uni-koeln.de}
\address[cologne]{Institute for Theoretical Physics, University of
  Cologne, 50937 K\"oln, Germany} 
\cortext[cor1]{Corresponding author}




\begin{abstract}
  The Braess paradox can be observed in road networks used by selfish
  users. It describes the counterintuitive situation in which adding a
  new, per se faster, origin-destination connection to a road network
  results in increased travel times for all network users. We study
  the network as originally proposed by Braess but introduce
  microscopic particle dynamics based on the totally asymmetric
  exclusion processes. In contrast to our previous
  work~\cite{PhysRevE.94.062312}, where routes were chosen randomly
  according to turning rates, here we study the case of drivers with
  fixed route choices. We find that travel time reduction due to the
  new road only happens at really low densities and Braess'
  paradox dominates the largest part of the phase
  diagram. Furthermore, the domain wall phase observed
  in~\cite{PhysRevE.94.062312} vanishes. In the present model gridlock
  states are observed in a large part of phase space. We conclude that
  the construcion of a new road can often be very critical and should
  be considered carefully.
\end{abstract}


\end{frontmatter}

\section{Introduction}

Urbanization is one of the big challenges of modern times. As the
world population grows, more and more people are moving into
cities~\cite{unpopulation}. With the growing population sizes, also
the transportation network has to adapt. The expansion of the city and
its transportation network, an interplay of top-down planning and
self-organizational processes~\cite{barthelemy2013self}, has to be
considered carefully to be efficient. The Braess paradox was
discovered by D.~Braess in 1968~\cite{Braess68,BraessNW05}. He
proposed a specific road network in which adding a new road
counterintuitively leads to higher travel times for all the network
users given that they minimize their own travel times selfishly. The
network consists of four individual roads forming two possible routes
from start to finish. Then a new road is added resulting in a new per
se faster route from start to finish\footnote{"Faster" in the sense
  of smaller travel time for a single particle compared to the original
  routes.}. A state of the system is characterized by the distribution of the vehicles onto the available roads. If a certain amount of drivers\footnote{Throughout this
  article we use the terminology "driver", "vehicle", "particle"
synonymously.} wants to go from start to
finish and they all want to minimize their own travel times, the system is in a stable state if all used routes have the
same travel time which is shorter than the travel times of any unused
routes. This is the so-called user optimum~\cite{wardrop1952} or Nash
equilibrium of the system. Braess showed that for specific
combinations of travel time functions of the roads and the total number
of cars, the user optimum of the system with the new road has higher
travel times than that of the system without the new road. This
appears to be a paradox since one would assume that an additional 
route increases the capacity and thus leads to a decrease of travel times.

Many efforts have been made in understanding Braess' paradox in
more general terms. Indeed it was shown that its occurrence is very
prevalent in congested networks~\cite{Steinberg83}. The regions of its
occurrence in certain models were
determined~\cite{pasprincipio,Nagurney10} and it was also shown to
occur in certain real-world scenarios~\cite{Kolata}. Furthermore,
analogues of the paradox were e.g. found in mechanical
networks~\cite{PenchinaP03}, energy networks~\cite{Witthaut2012},
pedestrian dynamics~\cite{MICE:MICE12209} or thermodynamic
systems~\cite{Bhattacharyya}. In most studies on the paradox the
focus was on macroscopical mathematical models of car traffic in
which the roads are treated as uncorrelated. Travel times of the
roads are given by functions which are linear in the number of cars
using the roads. This lead to the discovery, a general understanding
and also the observation of the effect in the real world and sparked
the ongoing fascination with this counterintuitive
phenomenon. Nevertheless, linear travel time functions and the fact
that in those models there are no correlation effects between the
roads results in a rather unrealistic description of traffic in the
network.

It is important to gain a deeper understanding of the paradox in more
realistic scenarios since this is relevant for the design of new roads
in real road networks. Especially the effects of a more realistic
microscopic dynamics and also inter-road correlations like jamming
effects or conflicts at road junctions need to be understood. In a
recent article~\cite{PhysRevE.94.062312} we have studied Braess'
network where the dynamics on the edges is given totally asymmetric
exclusion processes (TASEPs). The TASEP is a simple cellular
automaton~\cite{BIP:BIP360060102} and is now renowned as the
paradigmatic model for single-lane traffic. It covers a lot of
effects which are not included in deterministic mathematical
models. In \cite{PhysRevE.94.062312} we studied the case where all
drivers are identical and choose their routes stochastically.  The
route choice is determined through the turning probability on the
junction sites. We found that the paradox occurs at
intermediate global densities
$0.1\lesssim\rho_{\text{global}}\lesssim 0.3$. A large part of the
phase diagram is dominated by the so-called fluctuation-dominated
regime in which no stable travel times can be measured due to domain
walls of fluctuating positions (i.e. traffic jams of fluctuating
lengths).

In the present paper we analyse the same network with TASEP dynamics
on the edges. Instead of a random route choice based on turning
probabilities at junctions the particles have fixed routes.  This
corresponds e.g. to the scenario of daily commuters who stick to their
'favourite' routes. In the present model stable travel times can be
found throughout the whole phase diagram. The fluctuation-dominated
regime which is a defining phase for the model we studied previously
disappears. Braess' paradox is found to be even more prevalent in the
present case as the system shows Braess-like behaviour in almost the
whole phase space for densities $\rho_{\text{global}}\gtrsim 0.1$.
This implies that the paradox is even more prominent and important in
networks of microscopic transport models than we concluded due to our
findings in~\cite{PhysRevE.94.062312}.


\section{Model definition}
\subsection{The totally asymmetric exclusion process}
\begin{figure}[h!]
  \centering
  \includegraphics{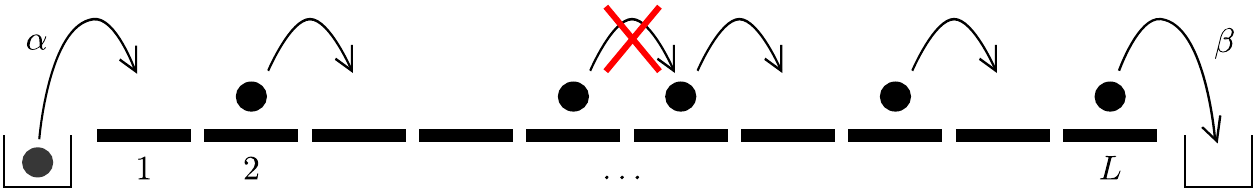}
  \caption{\label{fig:tasepeinfach} A single TASEP of length $L$ with
    open boundary conditions. In the case of random sequential update
    rules a site is chosen randomly. If it is occupied, the particle
    can jump to the next site if this next site is empty.}
\end{figure}

The TASEP is a one-dimensional paradigmatic stochastic transport
model. A single TASEP consists of $L$ cells ($L$ is also called the
length of a TASEP) which can be either empty or occupied by one
particle. In our analysis we use the so-called random sequential
update rules. With this update scheme, the dymanics works as follows:
With uniform probability one of the $L$ cells is chosen. If this cell
is occupied by a particle, the particle can jump to the next cell iff
the next cell is empty (see Figure~\ref{fig:tasepeinfach}). After $L$
of those updates, one timestep is completed\footnote{Note that not
necessarily all particles are updated in a timestep and some
particles can be updated more than once.}. In the case of open
boundary conditions, particles are fed onto site 1 from a reservoir
which is occupied with the entrance-probability $\alpha$. Particles
can leave the system from site $L$ with the exit probability
$\beta$. In the case of periodic boundary conditions site $L+1$ is
identified with site 1. Thus the total number of particles $M$ in
the system is constant and the system effectively becomes a ring.
In the stationary state of the periodic boundary case the density
profile is flat and the local density on each site $i$ equals the
global density $\rho(i)=\rho_{\text{global}}=M/L$. In this case, the
travel time, i.e. the number of timesteps a particle needs to complete one
round, is given by
\begin{equation}
 T=\frac{L}{1-\rho}. \label{eq:tperbc}
\end{equation}
The stationary state can also be determined analytically for open boundary
conditions~\cite{SchuetzD, DEHP, BlytheE07}. In this case,
Equation~(\ref{eq:tperbc}) is a good approximation for the travel times
for most combinations of $\alpha$ and $\beta$ if the density is
replaced by the average bulk density of the open
system~\cite{PhysRevE.94.062312}.


\subsection{Braess' network} \label{sec:braess-network}

\begin{figure}[h!]
  \centering
  \includegraphics{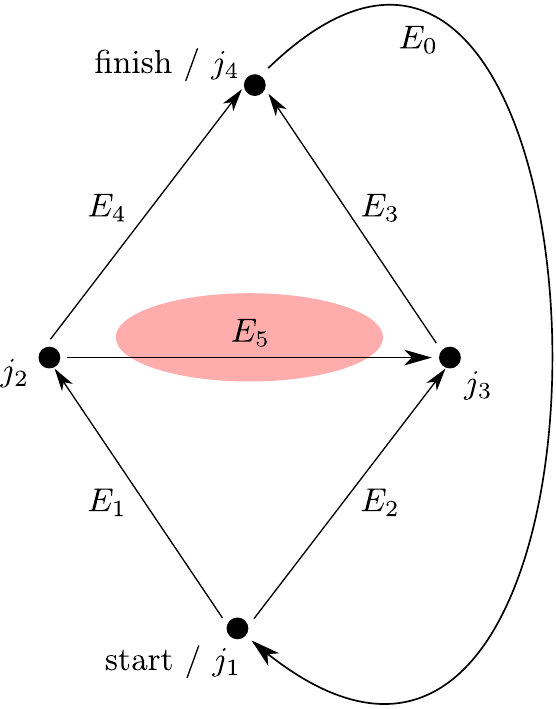}
  \caption{\label{fig:salinoperiodic} Schematic of Braess' network
    with the edges $E_0, \dots, E_5$ made up of TASEPs of length $L_i$
    joined through junction sites $j_1, \dots, j_4$. Edge $E_5$ is
    considered to be \textit{the new road} which is added to the
    system, resulting in a newly available route from start to
    finish. Through $E_0$ periodic boundary conditions are achieved
    and the total number of particles $M$ is kept constant. The
    lengths of the TASEPs are chosen according to
    Eqs.~(\ref{eq:lengths1}) and (\ref{eq:lengths2}) which
    according to Eqs.~(\ref{eq:l153-1})-(\ref{eq:l153-2}) results
    in route 153 being shorter than routes 14 and 23.}
\end{figure}

We consider the network shown in Figure~\ref{fig:salinoperiodic} which
is the network introduced by Braess in his original paper
\cite{Braess68,BraessNW05} but apply periodic boundary conditions.  We
examine the case that all particles want to go from the start at
junction site $j_1$ to the finish at junction site $j_4$. The edges
$E_i$ ($i = 0, \dots, 5$) of the network are TASEPs of lengths $L_i$
joined through junction sites $j_k$ ($k \in 1, \dots, 4$). The
junction sites behave as ordinary TASEP cells. They can take one
particle at a time. Periodic boundary conditions are achieved through
$E_0$ with $L_0=1$, coupling $j_4$ to $j_1$. Like this, the total
number of particles in the system $M$ and thus the global density
$\rho_{\text{global}}=M/(4+\sum_{i=0}^5L_i)$ is constant. Particles
that reach the finish point $j_4$ are fed back into the system via
$E_0$.

There are three different routes from start to finish. Route 14
leads from $j_1$ to $j_4$ via $E_1, j_2, E_4$.  Route 23 leads from
$j_1$ to $j_4$ via $E_2, j_3, E_3$. Edge $E_5$ is \textit{the new
  edge}, the newly build road which is added to the system, which
results in the third possible route, route 153, going from $j_1$ to
$j_4$ via $E_1, j_2, E_5, j_3, E_3$. Like in most previous work we
choose the system to be symmetric with
\begin{equation}
 L_1=L_3~~~~\text{and}~~~~L_2=L_4 \label{eq:lengths1}
\end{equation}
and $L_1<L_2$. The length of the new road is chosen as
\begin{equation}
 L_5\leq L_2-L_1-1,  \label{eq:lengths2}
\end{equation}
which leads to the new route 153 being shorter than routes 14 and 23:
\begin{eqnarray}
 \hat{L}_{153}&=5+L_1+L_3+L_5 \label{eq:l153-1}\\
 &\leq 4+L_1+L_2 \\
 &=\hat{L}_{14}=\hat{L}_{23}.\label{eq:l153-2}
\end{eqnarray}
Lengths of routes are denoted as $\hat{L}_i$.

In the following, the systems without and with $E_5$ are also refered
to as \textit{4link} and \textit{5link} system,
respectively. Corresponding variables are marked with the superscripts
'$(4)$' and '$(5)$'. Since we want to compare how the systems with and
without $E_5$ behave for the same demand $M$, we we introduce the two
different global densities for the 4link and the 5link system as
\begin{eqnarray}
  \rho^{(4)}_{\text{global}} &=& M/\left(4+\sum_{i=0}^4L_i\right) \label{eq:rho4}\\
  \rho^{(5)}_{\text{global}} &=& M/\left(4+\sum_{i=0}^5L_i\right)  \label{eq:rho5} \\
  &=& \frac{ 5+2L_{1}+2L_{2} }{ 2L_{2} + \frac{\hat{L}_{153}}{\hat{L}_{14}}(4+L_{1}+L_{2}) }\rho^{(4)}_{\text{global}}. \label{eq:rho4-5}
\end{eqnarray}
The relation between the two different global densities for the same
number of particles depends on $L_5$. The length $L_5$ can also be
given by the pathlength ratio $\hat{L}_{153}/ \hat{L}_{14}$, which
describes how long the new route is compared to the old ones.

Each particle has a fixed strategy (or personal route choice) which
does not change with time. Therefore the amounts of particles which
take specific routes are the relevant free parameters in our
system. For a fixed combination of the $L_i$ and $M$ the numbers of
particles which use the different routes 14, 23 and 153 are given by
$N_{14}$, $N_{23}$ and $N_{153}$. They are the tunable parameters in
the system subject to the condition
$N_{14}+N_{23}+N_{153}=M$. Specific choices of these numbers may drive
the system into specific states. Useful quantities are
\begin{eqnarray}
 n_{\text{l}}^{(j_1)}&=1-\frac{N_{23}}{M}, \label{eq:gammaeff}\\
 n_{\text{l}}^{(j_2)}&=\frac{N_{14}}{N_{14}+N_{153}},\label{eq:deltaeff}
\end{eqnarray}
i.e. the fraction of particles which turn 'left' on junctions $j_1$
and $j_2$, respectively. By varying $n_{\text{l}}^{(j_1)}$ and
$n_{\text{l}}^{(j_2)}$ from 0 to 1, all possible combinations of
$N_{14}$, $N_{23}$, $N_{153}$ and the corresponding values of certain
observables we are interested in can be visualized in simple
2d-color-plots.

Fixing the individual strategies is a major difference to our
previous work~\cite{PhysRevE.94.062312} on the Braess paradox. There
the particles did not have individual strategies, but were all equal:
particles sitting on junction sites $j_1$ or $j_2$ chose their routes
according to the turning probabilities $\gamma$ and $\delta$,
respectively. It is important not to confuse $n_{\text{l}}^{(j_1)}$
and $n_{\text{l}}^{(j_2)}$ with the turning probabilities $\gamma$ and
$\delta$ in our previous article. The $n_{\text{l}}^{(j_1/j_2)}$ are
not turning probabilities for jumping left but they describe the
fraction of particles jumping left on $j_1$/$j_2$ compared to the
total number of particles. In the present paper the level of
stochasticity is reduced to the random sequential update mechanism,
whereas in~\cite{PhysRevE.94.062312}, the turning probabilities were
an additional source of stochasticity. As will be seen in
Section~\ref{sec:results}, this difference leads to different
characteristics in the system - in our context especially in the
stability of measured travel time values as explained in detail for the
4link system in Section~\ref{sec:results_4link}. 

Networks comprised of TASEPs connected through junction sites and the
route choice process governed by turning probabilities received a lot
of research interest (see
e.g.~\cite{PhysRevLett.107.068702,PhysRevLett.110.098102,1367-2630-15-8-085005,ming2012asymmetric}). In
particular a network with a structure very similar to Braess'
network without $E_5$ was adressed in~\cite{PhysRevE.80.041128} in
some detail. The case of networks with particles choosing routes
according to personal strategies has to our knowledge not received as
much attention.

Both the model with turning probabilities and the present model with
fixed strategies can be regarded as realistic models of a commuter's
route choice scenario. Laboratory experiments in which humans
repetitively had to perform route choices were performed in similar
models with results suggesting that both fixed route choices or
turning probabilities could be
realistic. In~\cite{SELTEN2007394,rapoport2009choice} a network
similar to our network without $E_5$ and in~\cite{rapoport2009choice}
also the network with $E_5$ was examined. It turned out that in their
aim to minimize their individual travel times, in the network without
$E_5$ users kept varying their individual strategies while on average
the strategies stayed the same. This is an indication that the
model with turning probabilities is realistic. In the network with $E_5$,
strategy changes of individual users seemed to vanish after some time
rather indicating that the fixed-strategy model of the present paper
is realistic. Therefore both models seem to have some validity and
a mixture of both could be at play in reality.  Before addressing this
more complex scenario a good understanding of the differences 
between the two basic models seems to be helpful.


\subsection{Possible network states} 
\label{sec:possilbe-states} 

A \textit{state} of the road network is given by a certain
distribution of the particles onto the different routes, i.e. a
certain combination of $N_{14}$, $N_{23}$ and $N_{153}$. A road
network with selfish users is said to be in a \textit{stable state} if
the particles are distributed such that all used routes have the same
travel time which is lower than that of any unused routes. Such a
state is stable because it would not make sense for any particle to
change its route since a potential change would result in an increase
of the particle's travel time. If drivers have knowledge of the
current travel times of all routes, there is no incentive to alter
their current route choice in a stable state. This state is
also called the \textit{user optimum} ($uo$)~\cite{wardrop1952}. The
user optimum is to be distinguished from the \textit{system optimum}
($so$) which is the best state that can be reached from a global
perspective.  In the
following we define the system optimum as the state which minimizes
the maximum travel time in the system. This agrees with Braess' choice
in \cite{Braess68,BraessNW05}. Note that throughout
the literature other definitions of the system
optimum have been used, e.g. as the state minimizing the total travel
time~\cite{Thunig2016946}. 

\begin{figure}[h!]
  \centering
  \includegraphics{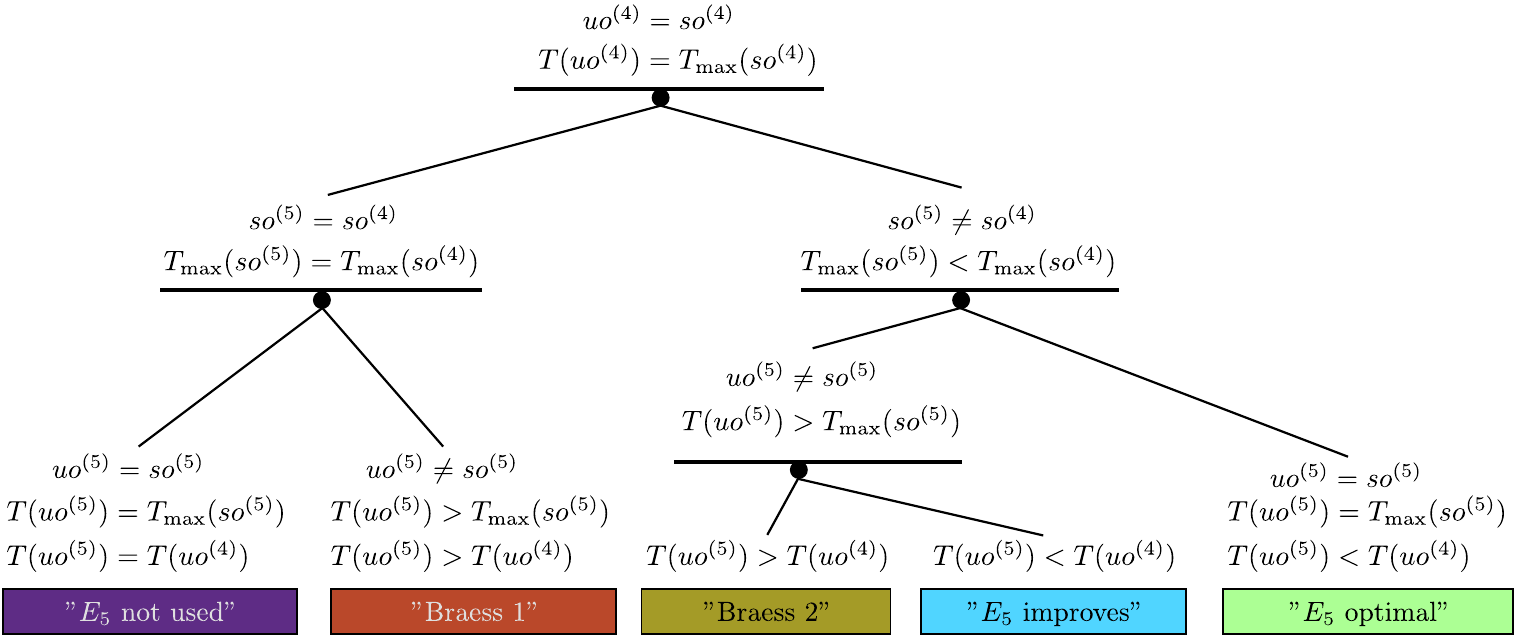}
  \caption{\label{fig:possiblestates}The possible states which can
    occur in our network. If the system optima of the 4link and the
    5link system are equal the system performance cannot be improved
    due to $E_5$, it is either unchanged ("$E_5$ not used") or
    rendered worse ("Braess 1"). If the system optima are unequal,
    then the 5link system optimum necessarily has a lower travel time
    than the 4link system optimum. Here, the system perfomance can be
    rendered worse ("Braess 2") or be improved ("$E_5$ improves" and
    "$E_5$ optimal").}
\end{figure}

The system optimum is not necessarily stable for the case of selfish
drivers~\cite{wardrop1952}. In our analysis we want to compare the
travel times in the user optima of the system without and with the new
edge $E_5$. Generally, the system's performance is improved by the
addition of $E_5$ if the travel time in the user optimum of the system
with $E_5$, i.e. the 5link system, is lower than that of the user
optimum in the system without $E_5$, i.e. the 4link system. Braess'
paradox occurs, i.e. the system's performance is decreased due to the
new road, if the travel times in the user optimum of the system with
$E_5$ are higher than the travel times in the user optimum of the
system without $E_5$~\cite{Braess68, BraessNW05}. To gain a deeper
understanding of the influence of $E_5$ on the system - beyond the
question of the occurrence of Braess' paradox - one can also compare
the maximum travel times in the system optima of the 4link and the
5link system. The possible states are summarized in
Figure~\ref{fig:possiblestates}.

If the system optima are the same (left branch of the tree in
Fig.~\ref{fig:possiblestates}), the system cannot be improved, even
for the case of non-selfish drivers or with traffic guidance
systems. If selfish drivers lead the 5link system into its optimum
($uo^{(5)}=so^{(5)}$) the new road will not be used at all ("$E_5$ not
used"). If this is not the case, the new road will be used, but the
travel times will increase ("Braess 1"). If the system optima of the
4link and the 5link are not the same, the system can potentially be
improved (right branch of the tree in
Fig.~\ref{fig:possiblestates}). Since the 4link system is included in
the 5link system, the maximum travel time of the system optimum in the
5link system can only be smaller than that of the 4link system
($T_{\text{max}}(so^{(5)})<T_{\text{max}}(so^{(4)})$). If there is a
traffic guidance system, the system can always always be
improved. For selfish drivers and without such a system, if the 5link
system does not reach its optimal state ($uo^{(5)}\neq so^{(5)}$), the
system's travel times in the user optimum can either be increased
("Braess 2") or decreased ("$E_5$ improves") due to the addition of
$E_5$. If the 5link system reaches its optimum ($uo^{(5)}=so^{(5)}$),
the system is said to be in the "$E_5$ optimal" state.

The "$E_5$ improves" and "$E_5$ optimal" states are the only cases in which
the new route is useful for the case of selfish drivers. In the
"Braess 2" case, the new road can be useful, if the system is driven
into its system optimum by an external traffic guidance system but is
not useful if the network is used by selfish drivers. In the other two
phases, the new road has no effect or even renders the situation
worse.

For this distinction between the possible states (or phases) it is
essential to define the system optimum as the state that minimizes the
maximum travel time. For different definitions, as e.g. the state
maximizing the flow or the state minimizing the total travel time,
this classification scheme does not necessarily hold.

If the user optima and system optima of the systems with and without
$E_5$ can be found for all different length ratios
$\hat{L}_{153}/\hat{L}_{14}$ and all different global densities, the
full phase diagram can be constructed according to this classification
scheme.  In \cite{PhysRevE.94.062312} this was done for the same
network, but with turning probabilities instead of fixed amounts of
particles choosing the different routes. In that analysis, the phases
"Braess 1", "$E_5$ improves", "$E_5$ optimal" and an additional domain
wall phase were found.


\subsection{Gridlocks} \label{sec:gl}

As described in Section~\ref{sec:possilbe-states} a state of the
network is determined by strategy distributions as given by
$N_{14},~N_{23},~N_{153}$ or
$n_{\text{l}}^{(j_1)},~n_{\text{l}}^{(j_2)}$ (see
Eqs.~(\ref{eq:gammaeff}) and (\ref{eq:deltaeff})). For certain
combinations of $\hat{L}_{153}/ \hat{L}_{14}$,
$\rho^{(5)}_{\text{global}}$ and $N_{14},~N_{23},~N_{153}$ all sites
of one path or of multiple paths can become completely occupied. This
corresponds to a total gridlock of the whole system, since each route
shares the sites $j_1$, $j_4$ and $E_0$. If one of the paths is
gridlocked, the whole system is gridlocked. Once a gridlock developed
it cannot dissolve, i.e. becomes a stationary state of the
system. In an ergodic system (with finite edge lengths $E_i$) a possible
gridlock state will always be reached at some point of the time
evolution\footnote{Note that, depending on the initial state, the
time to reach the gridlock state can be extremely long.}.
In the following we describe under which circumstances
gridlocks can occur. 

In our simulations the system was always initialized such that
particles were placed randomly but already on routes according to
their strategies. E.g. a particle following strategy 14 could not be
placed on $E_2$. The following arguments are based on this
initialization strategy. When other strategies are used,
e.g. completely random initial position irrespective of the strategy,
more gridlocks can occur. A simple example for this are initial states
where all sites on route 153 are occupied by the random initialization
even if this is not possible according to the strategy distribution.

\paragraph*{Gridlock on route 14} 
For the occurrence of a gridlock on route 14 the following three
conditions must be met:
\begin{equation}
 (N_{14}\geq L_{4}+1)~~~\land~~~(N_{14}+N_{153}\geq L_{1}+L_{4}+2)~~~\land~~~(M\geq\hat{L}_{14}=L_{1}+L_{4}+4).
\end{equation}
The first condition $N_{14}\geq L_{4}+1$ is necessary since all sites
on $E_4$ can only be occupied by particles of strategy 14. For a
gridlock to occur, additionally to all sites on $E_4$, also junction
$j_2$ must be occupied by a particle of the same strategy. As well as
all sites of $E_4$ and $j_2$, also all sites on $E_1$ must be occupied
at the same time. This can be by particles of strategies 14 or
153. Also junction $j_1$ must be occupied by a particle that wants to
turn left, thus one of strategy 14 or 153. This is represented in the
second condition $N_{14}+N_{153}\geq L_{1}+L_{4}+2$. For a complete
gridlock of route 14, also sites $j_4$ and $E_0$ must be
occupied. They can be be occupied by particles of any strategy 14, 153
or 23, which is represented in the third condition
$M\geq\hat{L}_{14}$.

\paragraph*{Gridlock on route 23} For the occurrence of a gridlock on
route 23 the following three conditions must be met:
\begin{equation}
 (N_{23}\geq L_{2}+1)~~~\land~~~(N_{23}+N_{153}\geq L_{3}+L_{2}+2)~~~\land~~~(M\geq\hat{L}_{23}=L_{3}+L_{2}+4).
\end{equation}
The first condition $N_{23}\geq L_{2}+1$ is necessary since all sites
on $E_2$ can only be occupied by particles of strategy 23. For a
gridlock to occur, additionally to all sites on $E_2$, also junction
$j_1$ must be occupied by a particle of the same strategy. As well as
all sites of $E_2$ and $j_1$, also all sites on $E_3$ must be occupied
at the same time. This can be by particles of strategies 23 or
153. Also junction $j_3$ must be occupied by one of those
particles. This is represented in the second condition
$N_{23}+N_{153}\geq L_{3}+L_{2}+2$. For a complete gridlock of route
23, also sites $j_4$ and $E_0$ must be occupied. They can be be
occupied by particles of any strategy 14, 153 or 23, which is
represented in the third condition $M\geq\hat{L}_{23}$.

\paragraph*{Gridlock on route 153} The conditions for the occurrence
of a gridlock on route 153 are a bit more complicated since there is
edge $E_5$ which can only be used by particles of strategy 153 and
there are the edges $E_1$ and $E_3$ which can be used by particles of
strategies 153 and 14 and 153 and 23, respectively. The first
condition which has to be met is
\begin{equation}
 N_{153}\geq L_5+1. \label{eq:gridlock153-1}
\end{equation}
This represents all sites of $E_5$ and junction $j_2$ being occupied
by particles of strategy 153. Then one has to consider the remaining
particles of strategy 153 which we denote by
$r_{153}=N_{153}-L_5-1$. They can now be distributed onto edges $E_1$
and $E_3$. As the second condition for a gridlock possibility on route
153, there has to exist an integer number $a\in \mathbb{N}$ with
$0\leq a \leq r_{153}$ such that
\begin{equation}
 (r_{153}-a+N_{14}\geq L_{1}+1)~~~\land~~~(a+N_{23}\geq L_{3}+1).\label{eq:gridlock153-2}
\end{equation}
The first part means that all sites on $E_1$ and also junction $j_1$
must be occupied by particles of strategies 14 or 153. The second one
means that junction $j_3$ and all sites on $E_3$ must be occupied by
particles of strategies 153 or 23. The third condition is
\begin{equation}
 M\geq \hat{L}_{153}=L_{1}+L_5+L_3+5.\label{eq:gridlock153-3}
\end{equation}
This ensures that sites $j_4$ and $E_0$ are occupied (by particles of
any strategy 14, 23 or 153). Summarzing, for a gridlock on route 153
to be possible the conditions in (\ref{eq:gridlock153-1}) and
(\ref{eq:gridlock153-3}) have to be met and an integer number
$a\in[0,r_{153}]$ has to exist such that the two conditions in
(\ref{eq:gridlock153-2}) can be fulfilled.
\begin{figure}[h!]
  \centering
  \includegraphics[width=0.329\columnwidth]{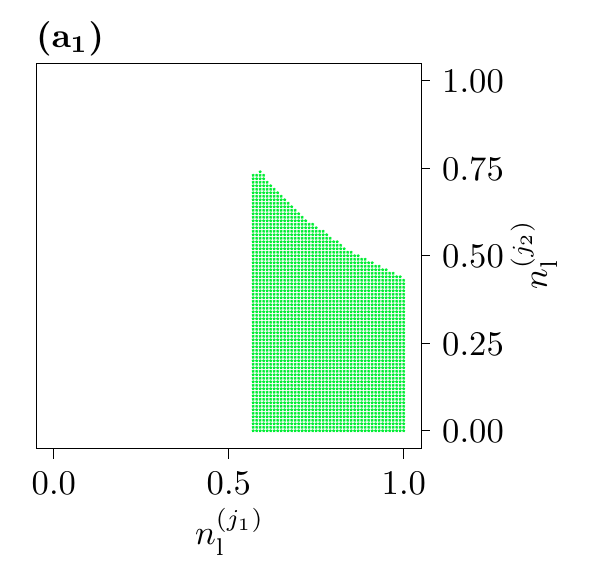}
   \includegraphics[width=0.329\columnwidth]{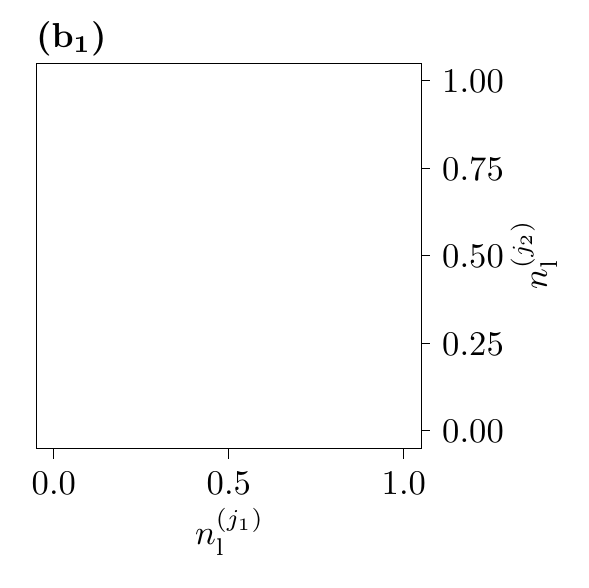}
    \includegraphics[width=0.329\columnwidth]{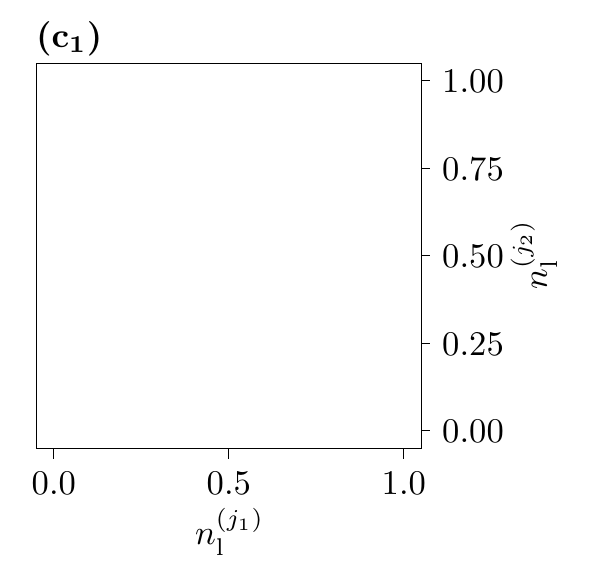}
  \includegraphics[width=0.329\columnwidth]{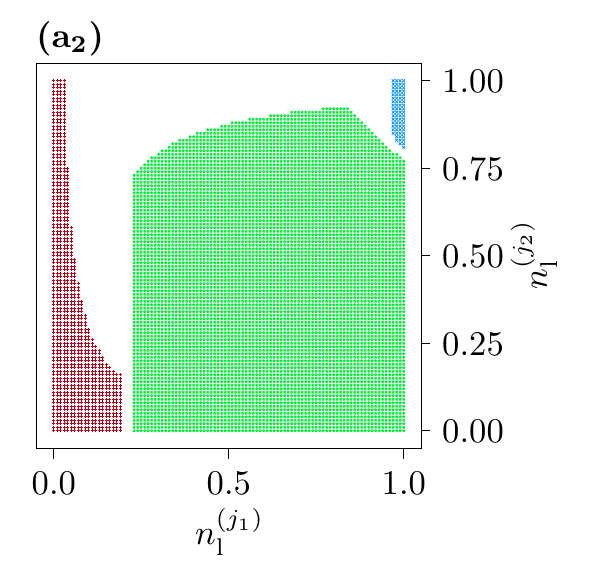}
   \includegraphics[width=0.329\columnwidth]{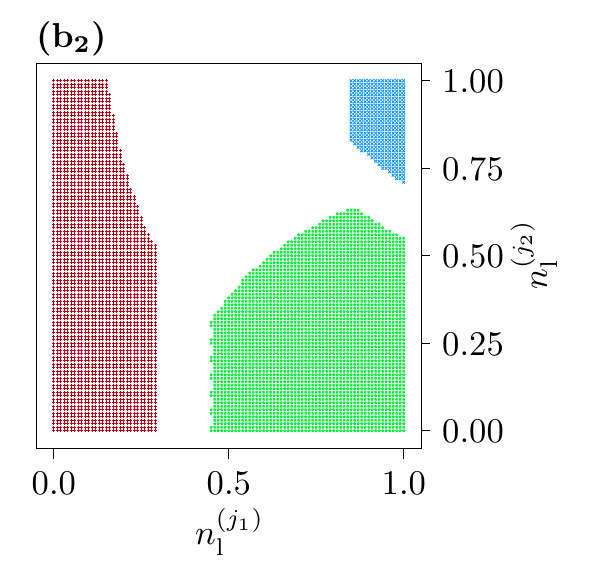}
    \includegraphics[width=0.329\columnwidth]{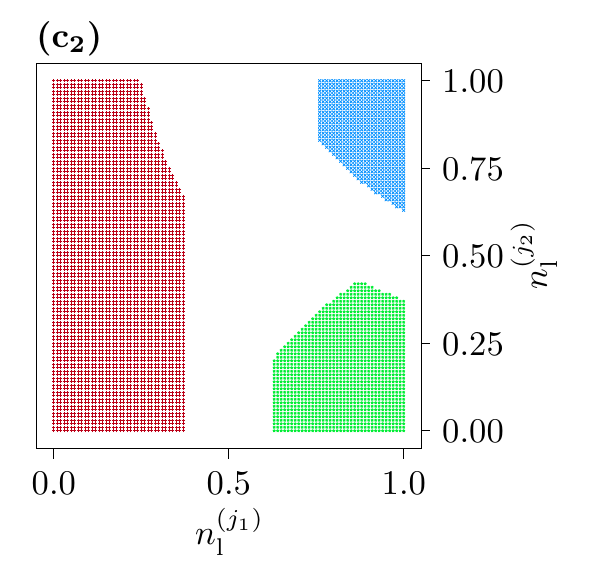}
  \includegraphics[width=0.329\columnwidth]{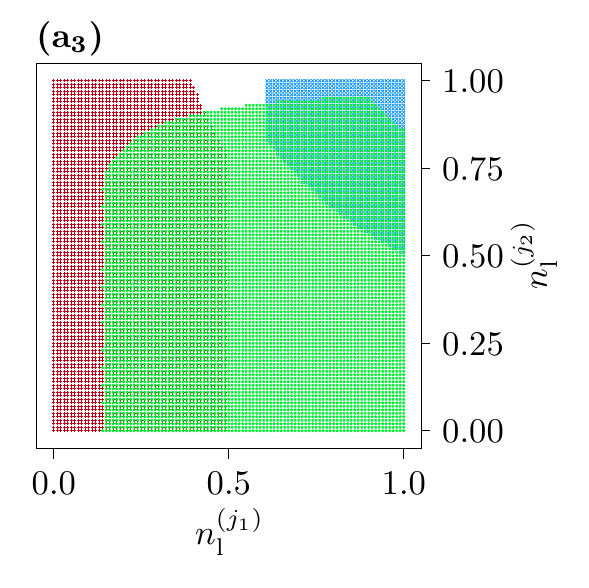}
   \includegraphics[width=0.329\columnwidth]{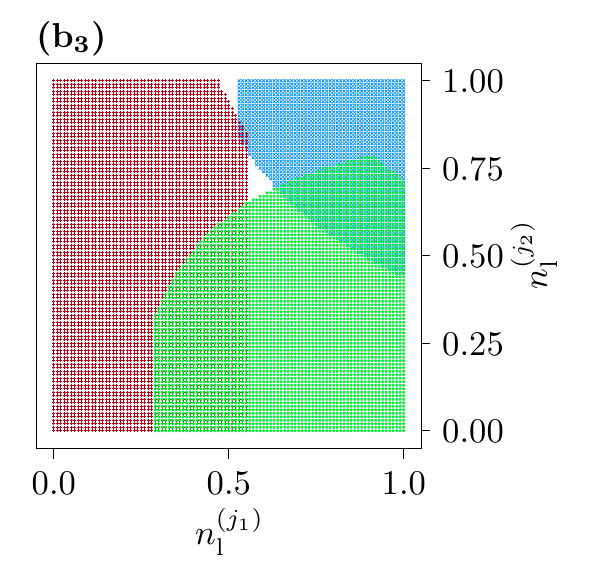}
    \includegraphics[width=0.329\columnwidth]{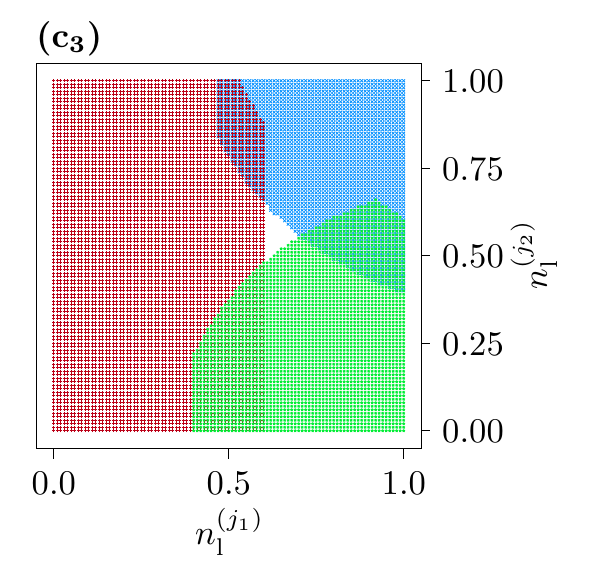}
    \caption{\label{fig:glproportions1}The
      $\left(n_{\text{l}}^{(j_1)},~n_{\text{l}}^{(j_2)}\right)$ -
      landscapes for $L_{1}=100$, $L_{2}=500$ and
      \textbf{(a}$\mathbf{_1}$\textbf{)}
      $\hat{L}_{153}/ \hat{L}_{14}=0.4$ and
      $\rho^{(5)}_{\text{global}}=0.2$,
      \textbf{(a}$\mathbf{_2}$\textbf{)}
      $\hat{L}_{153}/ \hat{L}_{14}=0.4$ and
      $\rho^{(5)}_{\text{global}}=0.5$,
      \textbf{(a}$\mathbf{_3}$\textbf{)}
      $\hat{L}_{153}/ \hat{L}_{14}=0.4$ and
      $\rho^{(5)}_{\text{global}}=0.8$,
      \textbf{(b}$\mathbf{_1}$\textbf{)}
      $\hat{L}_{153}/ \hat{L}_{14}=0.7$ and
      $\rho^{(5)}_{\text{global}}=0.2$,
      \textbf{(b}$\mathbf{_2}$\textbf{)}
      $\hat{L}_{153}/ \hat{L}_{14}=0.7$ and
      $\rho^{(5)}_{\text{global}}=0.5$,
      \textbf{(b}$\mathbf{_3}$\textbf{)}
      $\hat{L}_{153}/ \hat{L}_{14}=0.7$ and
      $\rho^{(5)}_{\text{global}}=0.8$,
      \textbf{(c}$\mathbf{_1}$\textbf{)}
      $\hat{L}_{153}/ \hat{L}_{14}=1.0$ and
      $\rho^{(5)}_{\text{global}}=0.2$,
      \textbf{(c}$\mathbf{_2}$\textbf{)}
      $\hat{L}_{153}/ \hat{L}_{14}=1.0$ and
      $\rho^{(5)}_{\text{global}}=0.5$,
      \textbf{(c}$\mathbf{_3}$\textbf{)}
      $\hat{L}_{153}/ \hat{L}_{14}=1.0$ and
      $\rho^{(5)}_{\text{global}}=0.8$. The landscapes were
      discretized in steps of 0.01. States with a gridlock possibility
      on routes 14, 23 and 153 are marked with
      \textcolor{blue-gl}{blue $\times$'s}, \textcolor{red-gl}{red
        $+$'s} and \textcolor{green-gl}{green $\star$'s}
      respectively.}
\end{figure}

Figure~\ref{fig:glproportions1} shows for which points in the
$\left(n_{\text{l}}^{(j_1)},~n_{\text{l}}^{(j_2)}\right)$ - landscape
(i.e. the phase space of the system) gridlocks on the routes are
possible for several combinations of $\hat{L}_{153}/ \hat{L}_{14}$ and
$\rho^{(5)}_{\text{global}}$. One can see that with growing global
density more and more strategies can become gridlocked. One can also
see that depending on the length ratio between the new route and the
old routes, different strategies can become gridlocked. For low values
of $\hat{L}_{153}/ \hat{L}_{14}$ more strategies with high
$n_{\text{l}}^{(j_1)}$ and low $n_{\text{l}}^{(j_2)}$ lead to a
gridlock on route 153 which makes sense since the new route is much
shorter compared to the old ones. For longer $L_5$, thus higher
$\hat{L}_{153}/ \hat{L}_{14}$, routes 14 and 23 are becoming more and
more likely to be gridlocked as well.
\begin{figure}[h!]
  \centering
  \includegraphics{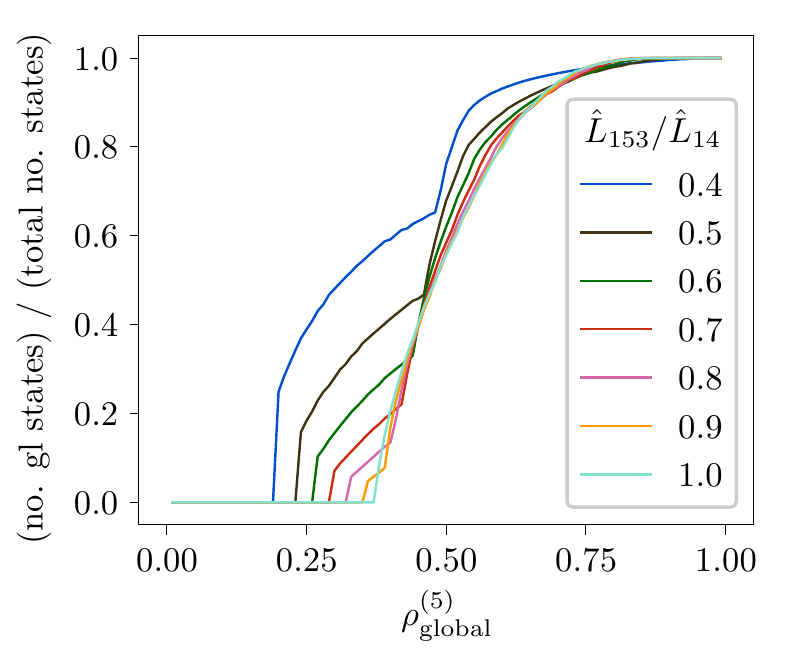}
  \caption{\label{fig:glproportions2}The number of states with
    gridlock probabilities over the total number of states against the
    global density $\rho^{(5)}_{\text{global}}$ for different values
    of $\hat{L}_{153}/ \hat{L}_{14}$. To calculate the ratio the
    $\left(n_{\text{l}}^{(j_1)},~n_{\text{l}}^{(j_2)}\right)$ -
    landscapes were discretized in steps of 0.01 and the number of
    states with gridlock possibility were counted and then divided by
    the total number of states.}
\end{figure}

In Figure~\ref{fig:glproportions2} the ratio of the number of states
with gridlock possibility over the total number of states is shown
against the global density for various pathlength ratios for
$L_{1}=100$ and $L_{2}=500$. To obtain this plot, the
$\left(n_{\text{l}}^{(j_1)},~n_{\text{l}}^{(j_2)}\right)$ - landscapes
were discretized in steps of 0.01 and the number of states with
gridlock possibility were counted. Since in our case $\hat{L}_{153}$
is always smaller than $\hat{L}_{14}=\hat{L}_{23}$, the lowest density
with a gridlock probabiliy is always the density with
$M=\hat{L}_{153}$ (cf. Eq.~(\ref{eq:gridlock153-3})). One can see that
for lower values of $L_5$ a large region of the phase space is
comprised of states with gridlock probabilities even for low densities
(e.g. over 60\% of states can become gridlocked at densities of
$\rho^{(5)}_{\text{global}}\approx 0.4$ for
$\hat{L}_{153}/ \hat{L}_{14}=0.4$). For longer $E_5$ less states can
become gridlocked. Please note that this figure depends on how fine
the discretization of space is chosen and also on how phase space is
desribed. It might look different if the phase space is
chosen to be three-dimensional and described by
$(N_{14},~N_{23},~N_{153})$ instead of
$\left(n_{\text{l}}^{(j_1)},~n_{\text{l}}^{(j_2)}\right)$. When a
gridlock state will occur in the system may depend on the
individual realization of the stochastic process. A closer look at
when gridlocked states occur is given in Section~\ref{sec:hatched}.


\subsection{Monte-Carlo observables} 
\label{sec:mc-observables} 

We analysed the system by employing Monte Carlo (MC)
simulations. Details of the measurement processes can be found
in~\ref{sec:app-measurement-process}. In a first step, the system and
user optima were found by analysing the values of the two observables
\begin{eqnarray}
\Delta T&=&|T_{14}-T_{23}|+|T_{14}-T_{153}|+|T_{23}-T_{153}|, 
\label{eq:deltat} \\
 T_{\text{max}}&=&\text{max}[T_i, i\in \{14,23,153\} ]. \label{eq:tmax}
\end{eqnarray}
$T_i$ denotes the travel time on route $i$, i.e. the time a particle
needs to go from $j_1$ to $j_4$ via this route. The user optima and
system optima are given by the combinations of $N_{14}$, $N_{23}$,
$N_{153}$ (or $n_{\text{l}}^{(j_1)}$ and $n_{\text{l}}^{(j_2)}$) which
minimize $\Delta T$ and $T_{\text{max}}$, respectively.

A true user optimum is characterized by
$\Delta T(n_{\text{l}}^{(j_1)}, n_{\text{l}}^{(j_2)})=0$ since then
all routes have the same travel times. If there are any unused routes
and the travel times of these unused routes are higher than that of the
used ones, $\Delta T$ is reduced to only the absolute values of the
travel time difference of the used routes (see~\ref{sec:app-special}
for an example of this case). In our simulations, a strategy with a
$\Delta T$ value close to zero is regarded as a user optimum since
the exact zero case is seldom found when measuring travel times.

The system optimum is given by the strategy which minimizes
$T_{\text{max}}$. By comparing the positions and travel time values of
the user and system optima of the 4link and 5link systems, the phase
diagram can be constructed according to the scheme described in
Section~\ref{sec:possilbe-states}. If a true user optimum is found,
all three routes have the same travel time. In this case this travel
time can be compared to the maximum travel time of the system
optimum. In cases where we could not find exact user optima, we
compared the maximum travel time of the 'closest candidate' for the
user optimum (state with lowest $\Delta T$) to the maximum travel time
of the system optimum. More details on the cases where no real user
optima could be found are given in Section~\ref{sec:hatched}.


\section{Results} \label{sec:results}

\subsection{Travel times in the 4link network} 
\label{sec:results_4link}

\begin{figure}[h!]
  \centering
  \includegraphics{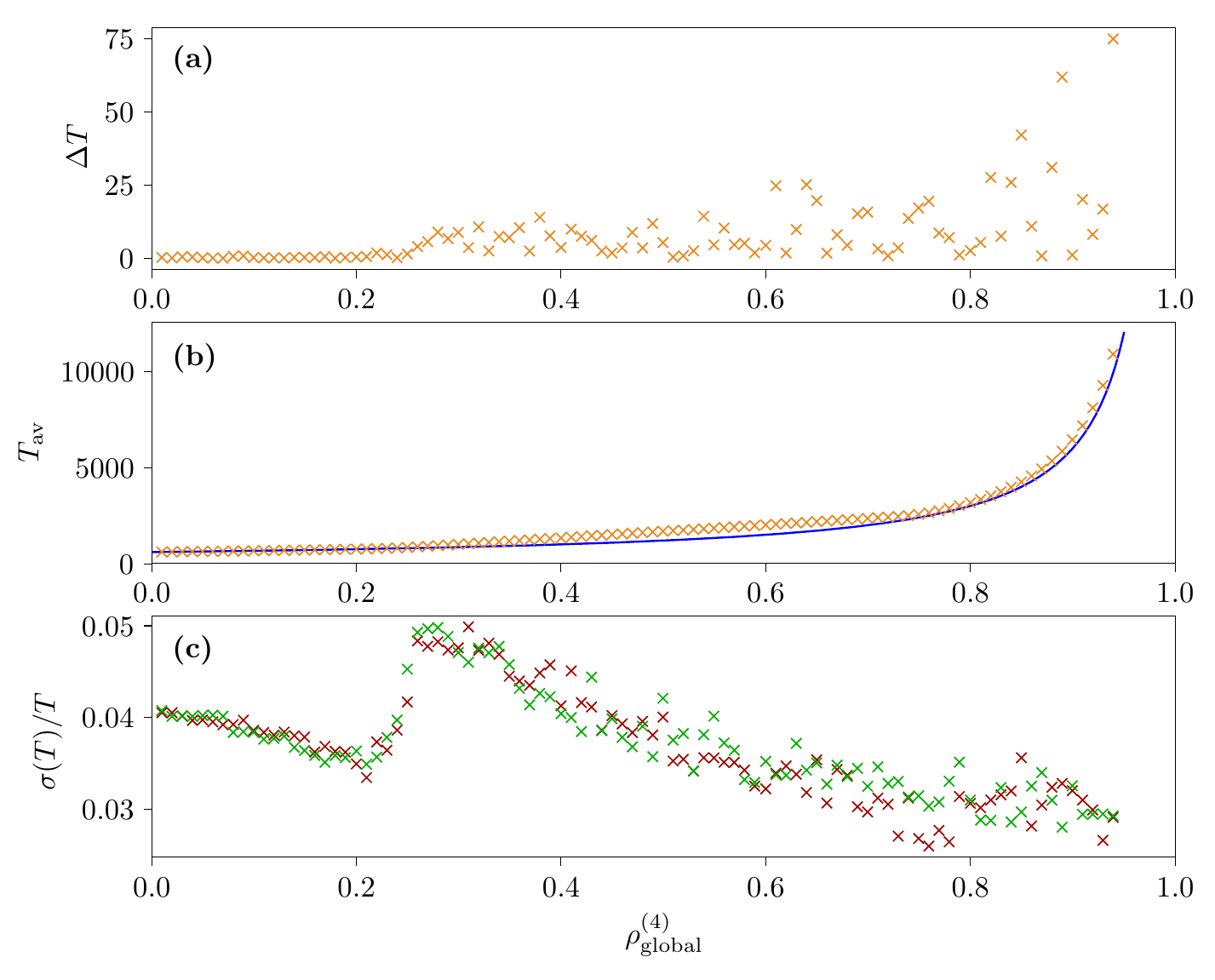}
  \caption{\label{fig:t4link}Results of MC simulations of the 4link
    system for 50\% of the particles choosing route 14 and 50\% choosing route 23 for the whole density regime
    $0\leq \rho_{\text{global}}^{(4)} \leq 1$ and $L_1=L_3=100$ and
    $L_2=L_4=500$. Part \textbf{(a)} shows, that throughout all
    densities, the travel times on routes 14 and 23 are almost
    equal. Part \textbf{(b)} shows the average travel times on these
    routes (\textcolor{orange-4link}{orange $\times$'s}) and for
    comparison the travel time of a single TASEP with $M/2$ particles
    according to Equation~(\ref{eq:tperbc})
    (\textcolor{blue-4link}{blue line}). One can see, that jamming
    at $j_4$ plays an important role for densities
    $\rho_{\text{global}}^{(4)}\gtrsim 0.2$. As can be seen in part
    \textbf{(c)}, the relative standard deviations of the travel times
    on both routes are below $5\%$ for all densities (values for route
    14/23 in
    \textcolor{red-4link}{red}/\textcolor{green-4link}{green}.}
\end{figure}

Since the 4link system, our reference system, is symmetric one expects
the user optimum and the system optimum to be given by half the
particles taking route 14 and the other half taking route 23 for all
possible global densities $\rho_{\text{global}}^{(4)}$. In
Fig.~\ref{fig:t4link} we can see that this is indeed the case.
In Figure~\ref{fig:t4link}~\textbf{(a)} we see the value of
$\Delta T=|T_{14}-T_{23}|$ (Equation~(\ref{eq:deltat}) reduces to this
form in the 4link system) against the global density. The value is
close to zero for all global densities which means that the travel
times are (almost) equal on both routes and this symmetric
distribution of the particles is indeed the user optimum. Since the
network is symmetric, this is also the system optimum, as any
unequal distribution of the particles would lead to a higher travel
time on the route with more particles. 

In Figure~\ref{fig:t4link}\textbf{(b)} the average of the travel times
measured on routes 14 and 23 ($T_{av}=(T_{14}+T_{23})/2$) is
shown. For comparison also the travel time of a single TASEP used by
$M/2$ particles (obeying Equation~(\ref{eq:tperbc})) is indicated by
the blue line. One can see that for densities
$\rho_{\text{global}}^{(4)}\gtrsim 0.2$ the jamming at $j_4$
plays an important role since the travel times are higher than those
of the single TASEP from this density upwards. 

\begin{figure}[h!]
  \centering
  \includegraphics{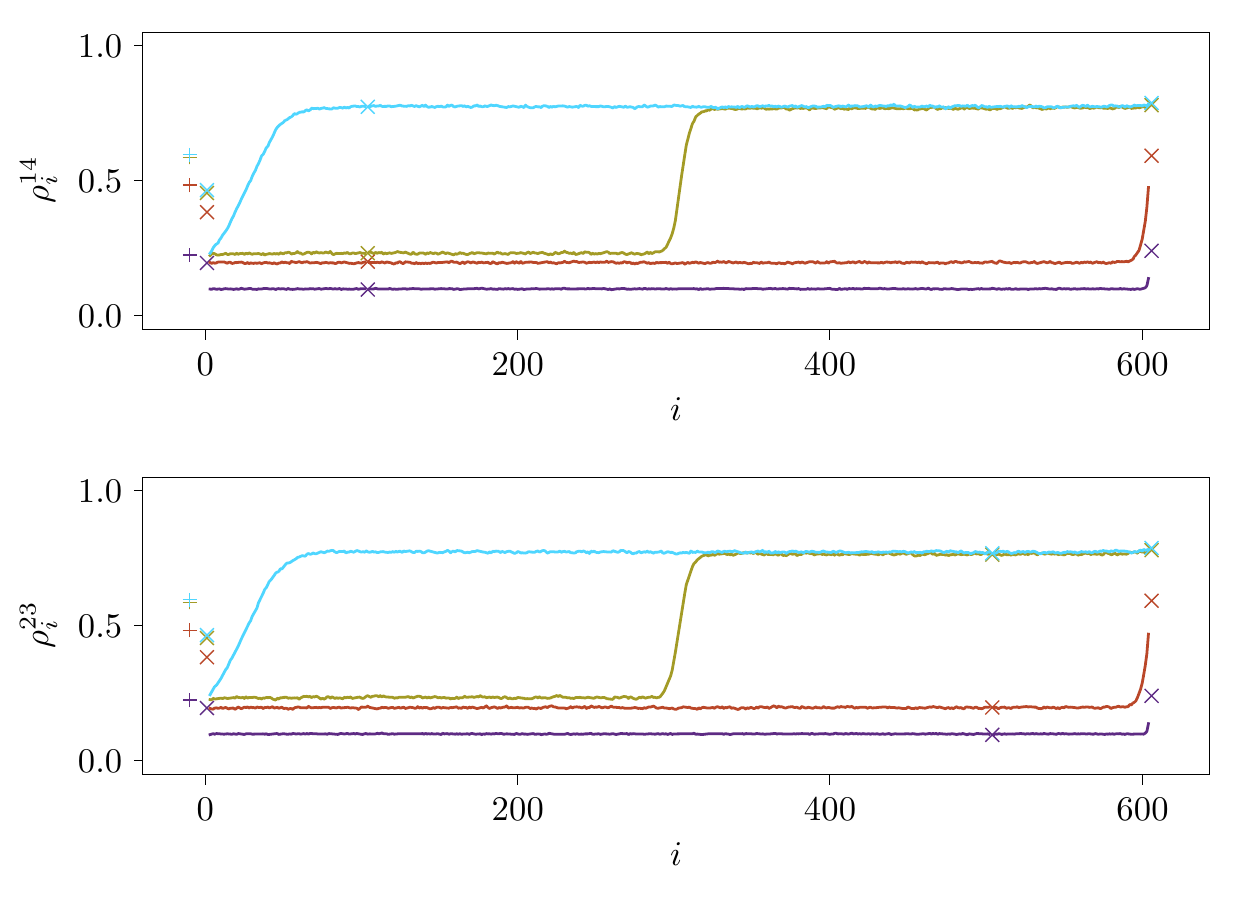}
  \caption{\label{fig:rho4link}Density profiles of the two routes in
    the 4link system with $L_1=L_3=100$ and $L_2=L_4=500$ for the four
    different global densities
    $\rho_{\text{global}}^{(4)} \in \{0.1,0.2,0.5,0.75\}$ printed in
    $\{$\textcolor{purple-density}{purple},
    \textcolor{orange-density}{orange},
    \textcolor{brown-density}{brown},
    \textcolor{blue-density}{blue}$\}$. The local densities
    $\rho_i^{14/23}$ on the two routes are shown against the position
    $i$. The denisity on $E_0$ is given by a $+$, the density of
    junction sites on the roads by $\times$'s. One can see that in all
    cases the density profiles are almost equal on both
    routes. Domains walls form at the same fixed positions on both
    routes.}
\end{figure}

In Figure~\ref{fig:t4link}\textbf{(c)} we can see the relative
standard deviation of the travel time measurements of both routes
\begin{equation}
\sigma(T)/T=\frac{1}{\bar{T}}\left(\frac{1}{n}\sum_{i=1}^n 
(T_i-\bar{T})^2\right)^{1/2}\label{eq:sd}
\end{equation}
with $n$ being the number of measurements and 
\begin{equation}
\bar{T}=T=\frac{1}{n}\sum_{i=1}^nT_i
\end{equation}
being the mean of all measured values. One can see that it stays below
$5\%$ for all densities. This means we measure stable travel time
values for all densities which is also supported by the measured
density profiles (Figure~\ref{fig:rho4link}). They show a sharp
domain wall between low and high density regions at a fixed position
which is the same on both routes.

\begin{figure}[h!]
  \centering
  \includegraphics{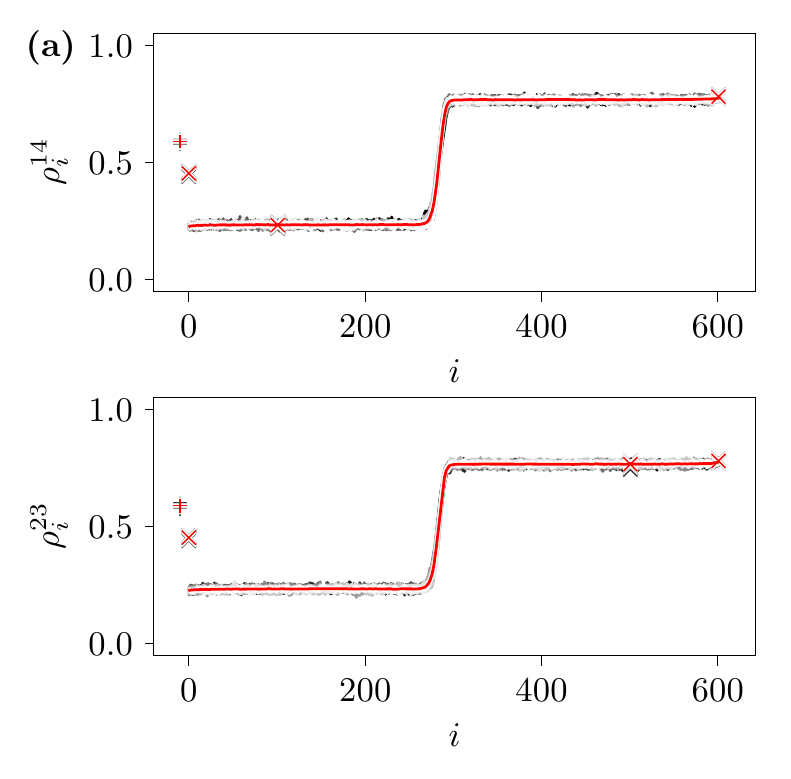}
  \includegraphics{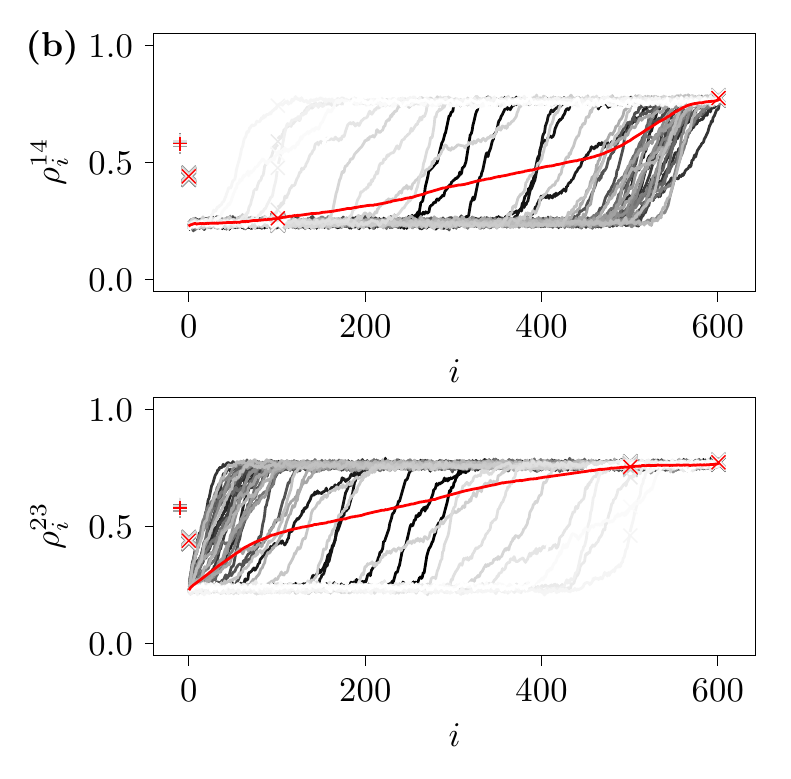}
  \caption{\label{fig:t4linkcomparison}A comparison of the density
    profiles of routes 14 and 23 in the 4link systems with
    $L_1=L_3=100$ and $L_2=L_4=500$ and $M=622$,
    $\rho_{\text{global}}^{(4)}\approx 0.52$ in the present less
    stochastic model (part \textbf{(a)}) and in the model with higher
    stochasticity (with turning probability $\gamma =0.5$) treated in
    our previous article~\cite{PhysRevE.94.062312} (part
    \textbf{(b)}). In both parts, the averaged density profiles over
    the whole measurement time of $10^6$ sweeps is shown in
    \textcolor{red}{red} and 50 different shortterm instances
    (averaged over $10^4$ sweeps each) are shown in 50 different
    shades of gray. The density on $E_0$ is given by a $+$, the
    density of junction sites on the roads by $\times$'s. One can see
    that in the present model, the domain wall position is fixed,
    while it fluctuates through the routes in a coupled manner in the
    model with higher stochasticity.}
\end{figure}

This is a very important difference to the model with higher
stochasticity due to turning
probabilities~\cite{PhysRevE.94.062312}. In the more stochastic case,
fluctuating domain walls dominate the density region
$0.29\lesssim\rho^{(4)}_{\text{global}}\lesssim0.75$ resulting in
(short-term) unstable travel time values. A comparison of the
different behaviours of the two models is shown in
Figure~\ref{fig:t4linkcomparison}.
In Fig.~\ref{fig:t4linkcomparison} \textbf{(a)} and \textbf{(b)}
density profiles of the present model and of that treated
in~\cite{PhysRevE.94.062312} are shown. The averaged density profiles
measured over the whole measurement time of $10^6$ sweeps are given in
red. In different shades of gray, different measurement instances
(averaged over $10^4$ sweeps each) are shown. As one can see in part
\textbf{(a)} of the figure, in the present model the domain wall is at
the same fixed position during the whole measurement process: all the
grey curves lie below the red curve. In the model with turning
probabilities this is not the case as seen in part \textbf{(b)}. Here,
the averaged density profile becomes close to a straight ascending line
on both paths. This is due to the fact that the
domain walls perform a coupled random walk on the two paths. If the
high density region gets longer on one path it gets shorter on the
other one respectively. The domain wall positions can be at any point
of the paths depending on the measurement time. Therefore, particles
entering the same route at different times may encounter a completely
different situation and thus have completely different travel times
even if the system is in its stationary state and none of the
parameters changes. This behaviour of the model with higher
stochasticity is treated in detail in~\cite{PhysRevE.94.062312}.

Due to the fluctuating domain walls and the resulting unstable travel
times in the model with higher stochasticity $uo$ and $so$ could not
be determined in a large intermediate global density region. Thus the
system's reaction to the addition of $E_5$ could not be classified
according to our scheme presented in Section~\ref{sec:possilbe-states}
in this density region. In the present case we obtain stable travel time
values and thus $uo$ and $so$ throughout the whole density region for
the 4link system. Thus we are able to compare the 5link system to
the 4link system in the whole density region.


\subsection{Phase diagram}

We used Monte Carlo simulations to obtain the user optima and system
optima of the 5link system for different combinations of $L_5$ and
$M$. States with potential gridlock formation as explained in
Section~\ref{sec:gl} were not considered as candidates for user or
system optima. Even if some of these states may, depending on the
initial configuration and realization of the stochastic process, have
values of $\Delta T$ close to zero in the beginning of the system's
time evolution, they all evolve into a gridlock steady state (more
details on this in Sec.~\ref{sec:hatched}). By comparing the travel
times of the found 5link user and system optima to those of the 4link
system's user and system optima for the same $M$ we could derive the
phase diagram of the system according to the classification shown in
Figure~\ref{fig:possiblestates}.  In~\ref{sec:app-measurement-process}
we describe our procedure to find the user and system optima in more
detail. The phase diagram (Fig.~\ref{fig:phases}) and the following
Fig.~\ref{fig:closerlook} show the influence of the control
parameters $L_5$ and $M$: The $x$-axis is always given by
$\hat{L}_{153}/\hat{L}_{14}$ which is the ratio of the lengths of the
new route 153 and the two old routes 14 and 23 (see
Eqs.~(\ref{eq:l153-1}) -- (\ref{eq:l153-2})). There are always two
$y$-axes which decode the number of particles $M$ via the global
densities in the 4link/5link systems
$\rho_{\mathrm{global}}^{(4)/(5)}$ (see Eqs.~(\ref{eq:rho4}) --
(\ref{eq:rho4-5})).


\subsubsection{Phase diagram's structure}
\begin{figure}[h!]
  \centering
  \includegraphics[width=0.49\columnwidth]{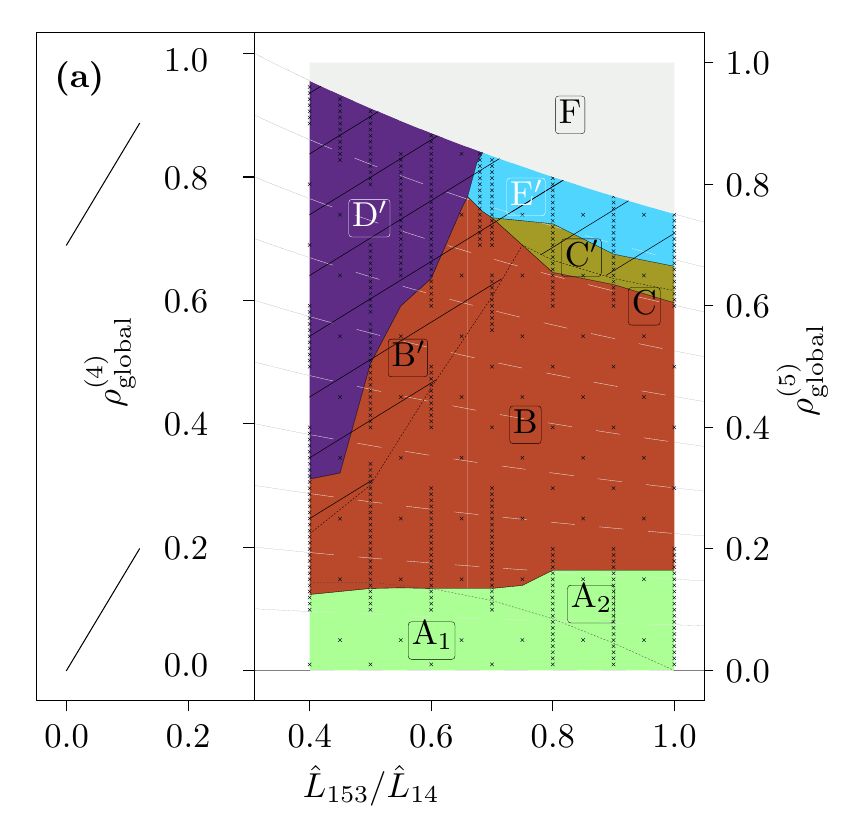}
  \includegraphics[width=0.49\columnwidth]{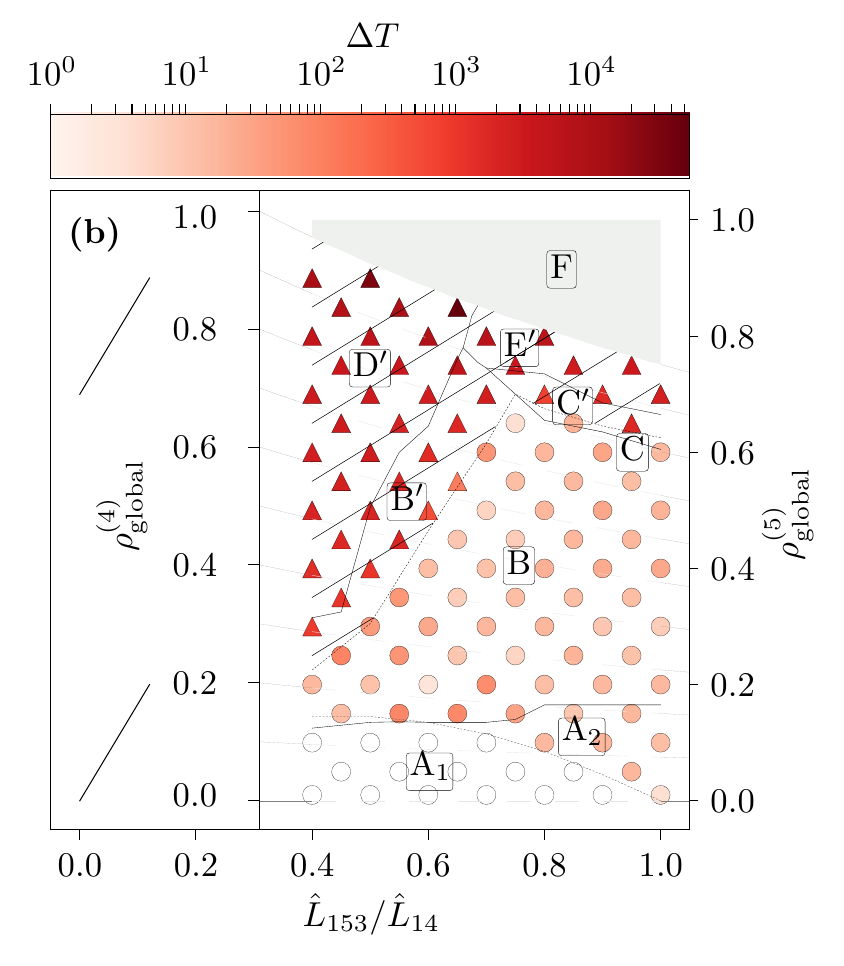}
  \caption{\label{fig:phases} The shown results were obtained for
    $L_1=L_3=100$, $L_2=L_4=500$ and varying lengths of $E_5$
    (resulting in the x-axis $\hat{L}_{153}/\hat{L}_{14}$) and $M$
    (resulting in the two y-axes
    $\rho_{\mathrm{global}}^{(4)/(5)}$). \textbf{(a)} The phase
    diagram according to the classification scheme given in
    Figure~\ref{fig:possiblestates}. The $\times$'s show where
    simulations were performed. The phase boundaries are drawn
    according to those simulations and are thus, due to this limited
    number of simulations, only a rough estimate of the phase
    boundaries. In phases A, B and C real user optima could by found
    while in the area marked by a hatching (phases B$^{\prime}$,
    C$^{\prime}$, D$^{\prime}$, E$^{\prime}$) no real user optima
    could be found. In phase F the 4link system is full. \textbf{(b)}
    The value of $\Delta T$ for selected measurement points. One can
    see that it is below 100 (indicated by coloured $\bigcirc$) in
    the unhatched area and above 100 (indicated by coloured
    $\bigtriangleup$) in the hatched area.}
\end{figure}

The phase diagram shown in
Figure~\ref{fig:phases}~\textbf{(a)} can be divided into three parts.
The trivial first part is called phase F. This phase is present for
really high densities in the 5link system. In this phase, there are
more particles in the 5link system than sites in the 4link system. The
4link system is thus completely blocked and the two systems cannot be
compared. One could argue that $E_5$ improves the system at these
high densities, but only in the sense that the 4link could not even
take up that many particles. 

The second part, consisting of phases A$_{1/2}$, B and C, is the part
in which real user optima of both the 4link and the 5link system exist
in the sense that states in which $\Delta T \approx 0$ exist,
i.e. states with almost equal travel times on all three routes.

The third part is given by the region above
the light dotted line which is marked by a hatching and consists of
the "primed" phases B$^{\prime}$, C$^{\prime}$, D$^{\prime}$ and
E$^{\prime}$. In all of these phases no real user optima exist in the
sense that no states with $\Delta T<100$ could be identified. This
means that in neither of these phases a distribution of the particles
onto the routes exists which leads to (almost) equal travel times on
all routes. This is a consequence of gridlocked states: Combinations
of $n_{\text{l}}^{(j_1)}$ and $n_{\text{l}}^{(j_2)}$ which would
potentially be a user optimum are not available since they would lead
to a gridlock.

The values of $\Delta T$ in some points of the phase diagram are shown
in Figure~\ref{fig:phases}~\textbf{(b)}. As can be seen, in
phases A, B and C the value is zero or of the magnitude of 10, while
it is higher than 100 in the hatched phase, reaching up to values of
the order of $10^4$ at really high densities.

Phases A$_{1/2}$ are "$E_5$ optimal" phases. In these phases, the
travel times in the $uo$ of the 5link system are lower than that of the
$uo$ in the 4link system and $uo^{(5)}=so^{(5)}$. In phase A$_1$, the
region below the dotted line at low path length ratios and low global
densities, the $uo$ of the 5link is given by all particles choosing
route 153. The new route is the only used route of the system. In
phase A$_2$, above the dotted line, all three routes are used in the
5link user optima.

Phase B is the "Braess 1" phase. Here the $so$ of the 5link is equal
to that of the 4link, meaning half of the particles choose route 14
and the other half route 23. The user optimum of the 5link is given by
another distribution of particles onto the three routes, such that all
three routes have the same \textit{higher} travel time. The "Braess 1"
phase actually dominates the largest part of the phase diagram. There
is also a very small part of the phase diagram at densities around
$\rho_{\mathrm{global}}^{(4)}\approx 0.8$ and
$\hat{L}_{153}/\hat{L}_{14}\gtrsim 0.7$, in which the "Braess 2" phase
(phase C) is found, meaning that with traffic regulations the system
could be improved due to $E_5$
($T_{\mathrm{max}}(so^{(5)})<T_{\mathrm{max}}(so^{(4)})$), but without
regulations - in the $uo$ - the travel times are increased due to the
addition of $E_5$.

In the hatched area, in the 5link system no real user optima with
equal travel times on all routes exist. This is because combinations
of $n_{\text{l}}^{(j_1)}$ and $n_{\text{l}}^{(j_2)}$ which could
potentially lead to equal travel times on all routes lead to gridlocks
in the system according to Section~\ref{sec:gl}. Only states without
the possibility of a gridlock were taken into account for the
construction of the phase diagram. In most cases, the state with the
lowest value of $\Delta T$ and without gridlock possibility is a state
in which the travel time on route 153 is lower than on the other
two. When more particles choose route 153, $T_{153}$ increases. But
before a state with equal travel times on all routes is reached, route
153 and subsequently all three routes come to a complete
gridlock. More details on the hatched area and an illustrative example
can be found in Section~\ref{sec:hatched}. 

For the construction of our phase diagram we used the state with the
lowest value of $\Delta T$ and without gridlock possibility we could
find and compared its maximum travel time to that of the system
optimum and the user optimum of the 4link system. It is important to
keep in mind that even though in this manner we could identify
different phases inside the hatched region, there are no real user
optima in the 5link in all of the primed phases. In a system with real
drivers more and more drivers would tend to switch routes with the
desire to reduce their own travel time and in the long run the system
will evolve into a complete gridlock.

Here, judging from the maximum travel times in the state with the
lowest $\Delta T$, we could identify two phases in which the system
could not be improved even if the traffic was externally
regulated. These are the "$E_5$ not used - like" phase (phase
A$^{\prime}$) and the a "Braess 1 - like" phase (phase
B$^{\prime}$). In both of those phases the $so$ of the 5link equals 
that of the 4link. While in the former this is also the state with
the lowest value of $\Delta T$ (and real drivers would thus tend to
switch to route 153), in the second one another state with a higher
maximum travel time gives the lowest value of $\Delta T$. In the
"Braess 2 - like" phase (phase C$^{\prime}$) the system could be
improved by the addition of $E_5$ if the traffic is regulated but is
not for selfish drivers. In the "$E_5$ improves - like" phase (phase
E$^{\prime}$), the maximum travel time in the state with the lowest
$\Delta T$ in the 5link is lower than that of the 4link but is not as
low as in the 5link's system optimum.

Summarizing we can conclude that in a system of selfish drivers
without traffic regulations the addition of $E_5$ only reduces travel
times at low densities $\rho_{\mathrm{global}}^{(5)}\lesssim 0.2$
(phases A$_{1/2}$). Here the system with the new road will also be in
its optimal state ($uo^{(5)}=so^{(5)}$). At really high densities, the
system shows "$E_5$ improves - like" behaviour (phase E$^{\prime}$),
meaning that in the 5link system the state in which all three travel
times are closest to each other has a lower maximum travel time than
the user optimum of the 4link system. Still the 5link is not in its
optimum in this state and, more importantly, this is not a real user
optimum. Instead the system is prone to gridlocks when used by real
drivers since more and more of them would tend to choose route 153. In
most parts of the system, the addition of $E_5$ leads to higher travel
times in the system used by selfish drivers (phases B, B$^{\prime}$,
C$^{\prime}$) or the new road will be ignored (phase D$^{\prime}$). In
most of these phases (phases B, B$^{\prime}$, D$^{\prime}$) even if
external traffic guidance was applied, $E_5$ would not reduce travel
times.


\subsubsection{The region without proper user optima} 
\label{sec:hatched}

\begin{figure}[h]
  \centering
  \includegraphics[width=0.49\columnwidth]{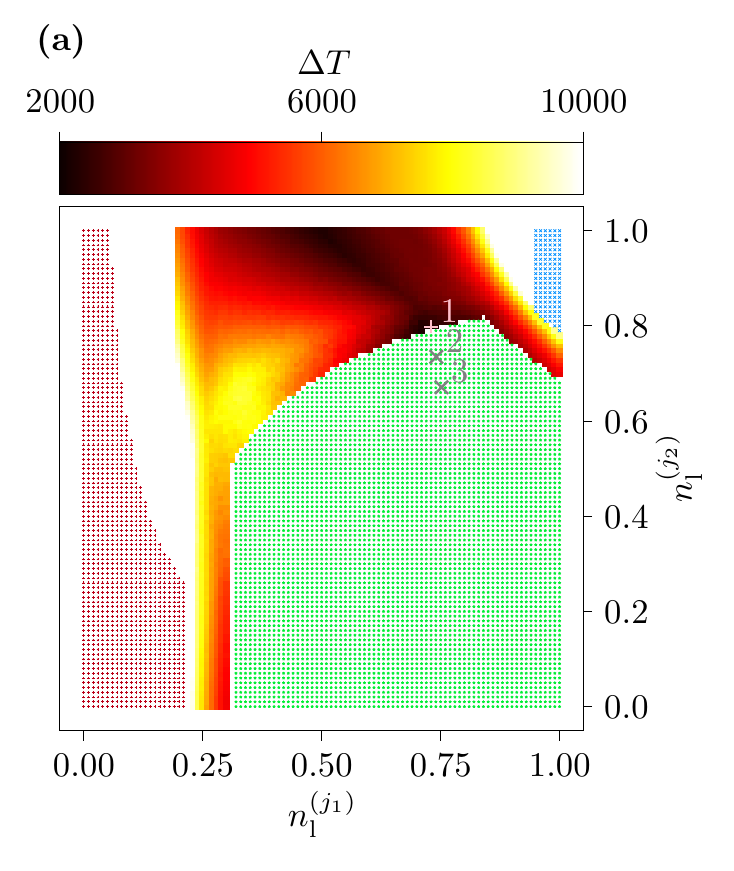}
  \includegraphics[width=0.49\columnwidth]{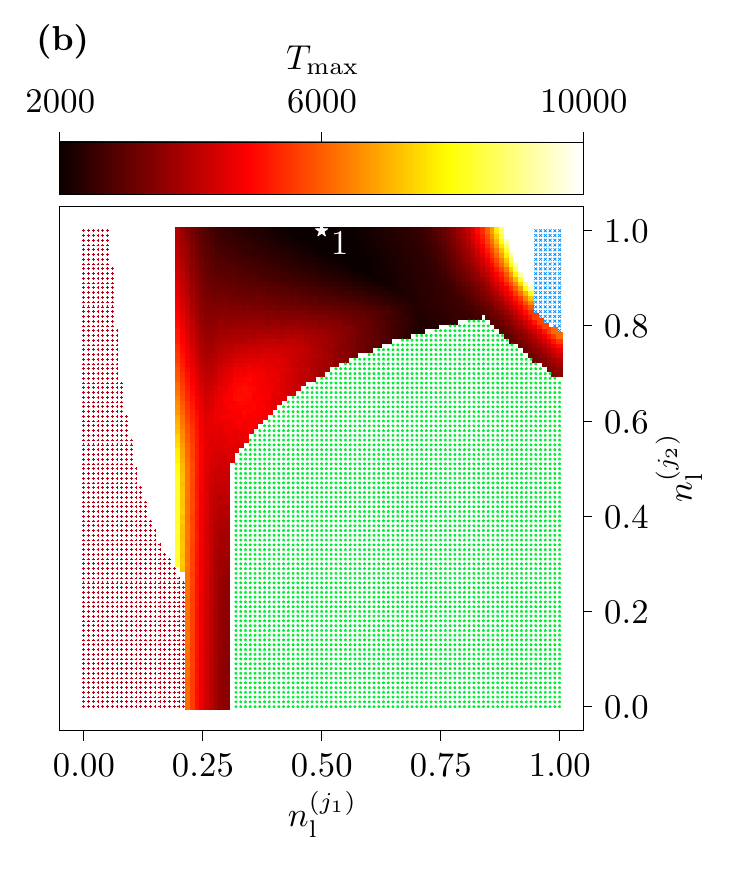}
  \caption{\label{fig:hatchedexample}The $n_{\text{l}}^{(j_1/j_2)}$ -
    landscapes of \textbf{(a)} the $\Delta T$ and \textbf{(b)} the
    $T_{\mathrm{max}}$ -values for $L_5=97$ and $M=638$, which means
    $\hat{L}_{153}/\hat{L}_{14}=0.5$ and
    $\rho_{\mathrm{global}}^{(5)}=0.49$. States with a gridlock
    possibility on routes 14, 23 and 153 are marked with
    \textcolor{blue-gl}{blue $\times$'s},
    \textcolor{red-gl}{red~$+$'s} and \textcolor{green-gl}{green
      $\star$'s} respectively. Simulations were performed for regions where no gridlock is possible and $n_{\text{l}}^{(j_1/j_2)}$ was
    sweeped in steps of 0.01. The pink point $+1$ is the point with
    the lowest value of $\Delta T$. Thus
    this is the point which is used for the phase diagram. The grey
    points $+2$ and $+3$ have smaller values of $\Delta T$ but route
    153 can gridlock. The white point $\star 1$ has the lowest value
    of $T_{\mathrm{max}}$ and is thus the system optimum.}
\end{figure}

Here we take a closer look at the phases in the area of the phase
diagram which is marked by a hatching (Figure~\ref{fig:phases}) by
analysing an exemplary point of this region: we look at the point with
the parameters $\hat{L}_{153}/\hat{L}_{14}=0.5$ and
$\rho_{\mathrm{global}}^{(5)}=0.49$. According to the phase
diagram shown in Fig.~\ref{fig:phases} this is a point of the B$^{\prime}$
phase ("Braess1 - like" phase). In Fig.~\ref{fig:hatchedexample} we
show the $n_{\text{l}}^{(j_1/j_2)}$ - landscapes and the corresponding
$\Delta T$ and $T_{\mathrm{max}}$ values. In
Fig.~\ref{fig:hatchedexample} certain points are marked and the
corresponding travel time values of the three paths and the $\Delta T$
and $T_{\mathrm{max}}$ values for these points are shown in
Table~\ref{tab:hatched-exp}.

\begin{figure}[h]
  \centering
  \includegraphics{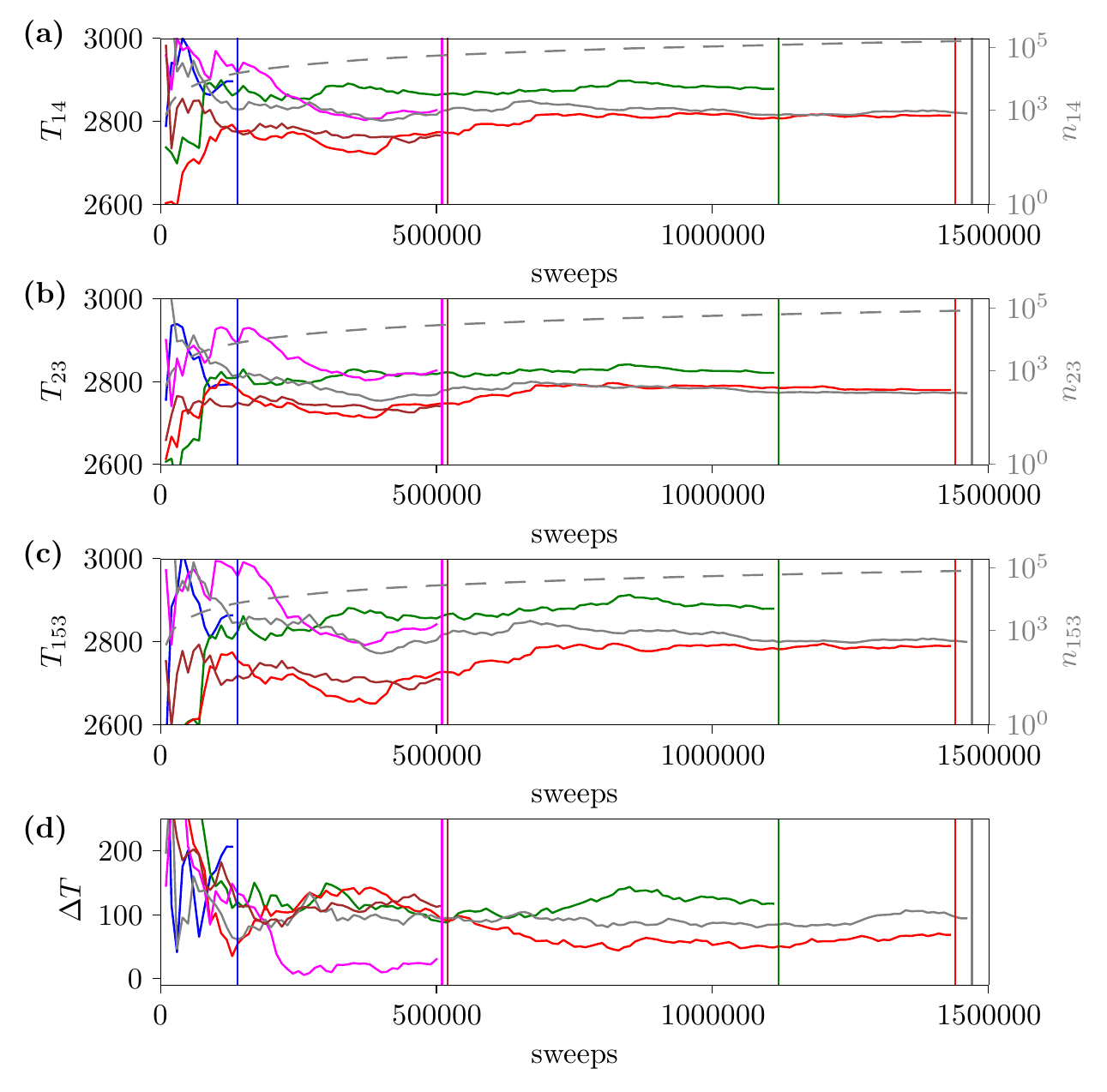}
  \caption{\label{fig:hatchedpoint3}The time evolution of \textbf{(a)}
    - \textbf{(c)} the measured mean values of the travel times of the
    three routes and \textbf{(d)} the value of $\Delta T$ for the
    state $\times 3$ shown in Figure~\ref{fig:hatchedexample}. The
    different coloured curves show six different instances of the
    system realized with six different seed values for the random
    number generator. The second y-axes on the right sides of the
    first three figures show how many times $n_i$ the values of $T_i$
    were measured, represented by the dotted grey line. The vertical
    lines show the points in time when the system gridlocked on route
    153.}
\end{figure}

In part \textbf{(b)} of Fig.~\ref{fig:hatchedexample} the point
$\star 1$ represents the system optimum of this parameter set. It is
given by half the particles choosing route 14 and the other half route
23. This point has the lowest value of $T_{\mathrm{max}}=1789$. It is
not the user optimum as it has a high value of $\Delta T=2230$
since the travel time on route 153 is much lower than on the other
routes. More and more particles would tend to switch to route 153. The
point $+1$ shown in Fig.~\ref{fig:hatchedexample}~\textbf{(a)} is
the point which we used for our construction of the phase diagram. It
is the point with the lowest value of $\Delta T=2215$ of all the
points without gridlock possibility. From Table~\ref{tab:hatched-exp}
we see that at this point route 153 still has a much lower travel time
than the routes 14 and 23. In a system with real selfish drivers, more
and more drivers would thus switch onto route 153.
\begin{table}[h]
  \caption{\label{tab:hatched-exp} The $n_{\text{l}}^{(j_1/j_2)}$,
    $\Delta T$, $T_{\mathrm{max}}$ and travel time values of the 
    routes for the 4 points which are marked in
    Figure~\ref{fig:hatchedexample}~\textbf{(a)} and
    ~\textbf{(b)}. Point $+ 1$ is the point with 
    the lowest value for $\Delta T$ without a gridlock possibility. 
    Thus this is the point we chose for construction of the phase
    diagram. Point $\star 1$ is the system optimum. The points $\times
    1$ and $\times 2$ are states with gridlock possibilities on 
    route 153. The measured travel time values are marked with a star 
    because they were measured before the system gridlocked and are 
    not stationary state values.}
\begin{center}
    \begin{tabular}{| c | l | l | l | l | l | l | l |}
     \hline
    Point &  $n_{\text{l}}^{(j_1)}$ & $n_{\text{l}}^{(j_2)}$ & $T_{14}$ & $T_{23}$ & $T_{153}$ &  $\Delta T$ & $T_{\mathrm{max}}$ \\ \hline
    $+ 1$ & 0.730 &  0.798 & 2270 & 2047 & 1162 & 2215 & 2270 \\ \hline
    $\times 2$ & 0.741 & 0.735 &  2113$^{\star}$ & 2015$^{\star}$ & 1539$^{\star}$ & 1148$^{\star}$ & 2113$^{\star}$ \\ \hline
    $\times 3$ & 0.752 & 0.671 & 2797$^{\star}$ & 2744$^{\star}$ & 2748$^{\star}$ & 106$^{\star}$ & 2797$^{\star}$  \\ \hline\hline
    $\star 1$ & 0.5 &  1.0 & 1789 & 1789 & 674 & 2230 & 1789 \\ \hline
    \end{tabular}
\end{center}
\end{table}
From Figure~\ref{fig:hatchedexample}~\textbf{(a)} we know that
if more particles choose route 153, a gridlock on that route becomes
possible. We marked two more points (point $\times 1$ and $\times 2$)
in this figure. From Table~\ref{tab:hatched-exp} we see that the value
of $\Delta T$ decreases for those two points. It is important to keep
in mind though that the travel time values of points $\times 1$ and
$\times 2$ were measured before the system gridlocked. In
Figure~\ref{fig:hatchedpoint3} we see how travel times of the
individual routes and the value of $\Delta T$ develop during the
measurement process at point $\times 3$.
For this figure six instances of the system with different seed values
for the random number generator were generated and the travel times
were measured during the evolution of the system. The system was not
relaxed before measurements begun (the system was always relaxed for
all measurements which were used for the phase diagram though -
compare~\ref{sec:app-measurement-process} for details on our general
measurement process). The relaxation was skipped here since otherwise
the system may have already gridlocked during the relaxation
process. One can see that the state $\times 3$ is actually a good
candidate for a user optimum since the value of $\Delta T$ seems to be
very low since all three routes have similar travel times (also
compare Table~\ref{tab:hatched-exp} for the numbers). All six
instances of the system gridlock at some point of the time
evolution. The earliest gridlock occurs after 130000 sweeps (blue
line) and the latest after 1470000 sweeps (grey line). This
example shows rather clearly that in the hatched area of the phase
diagram, if the system was actually used by intelligent selfish
particles, the gridlock possibility is very high. This is because
states with gridlock possibilities have actually lower possible values
of $\Delta T$. Thus more and more particles would tend to switch routes with the aim of reducing their own
travel times but with the risk of causing a gridlock of the whole
network. This is why for our phase diagram (Figure~\ref{fig:phases})
and for our further analysis on the influence of $E_5$ on the system (see
Sec.~\ref{sec:closerlook} and Fig.~\ref{fig:closerlook} therein)
we only considered points without gridlock possibility and measured
the travel time values of the steady states.


\subsubsection{A closer look at the influence of the new edge} 
\label{sec:closerlook}

\begin{figure}[h]
  \centering
  \includegraphics[width=0.49\columnwidth]{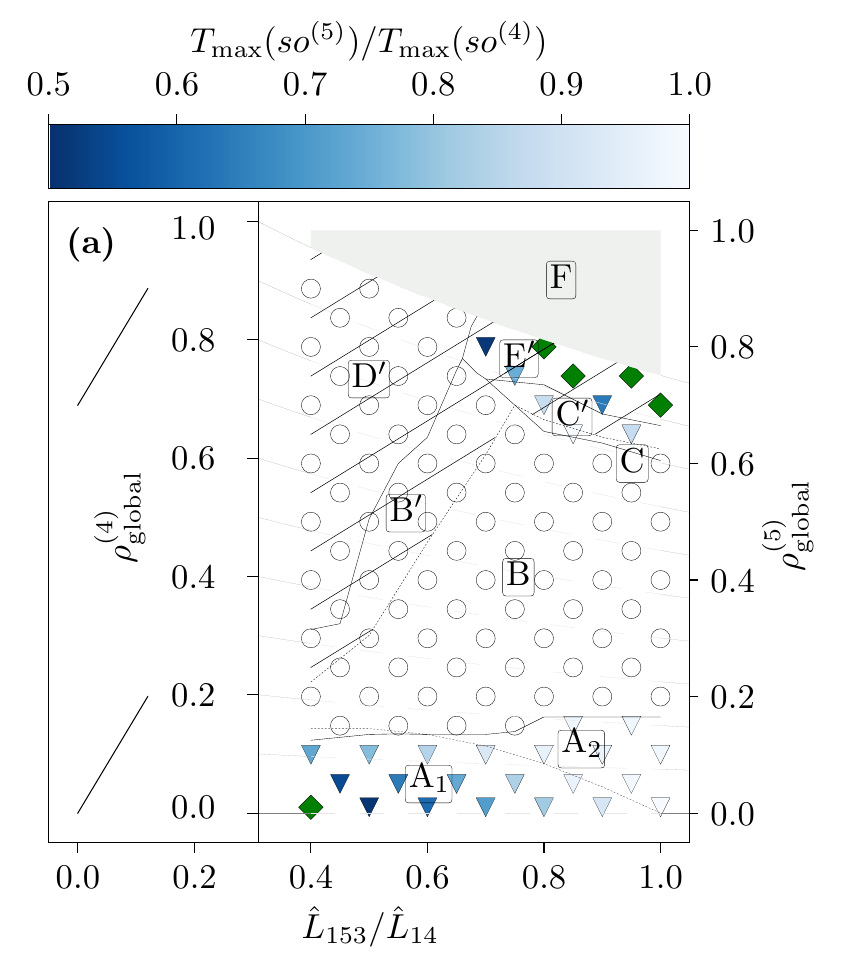}
  \includegraphics[width=0.49\columnwidth]{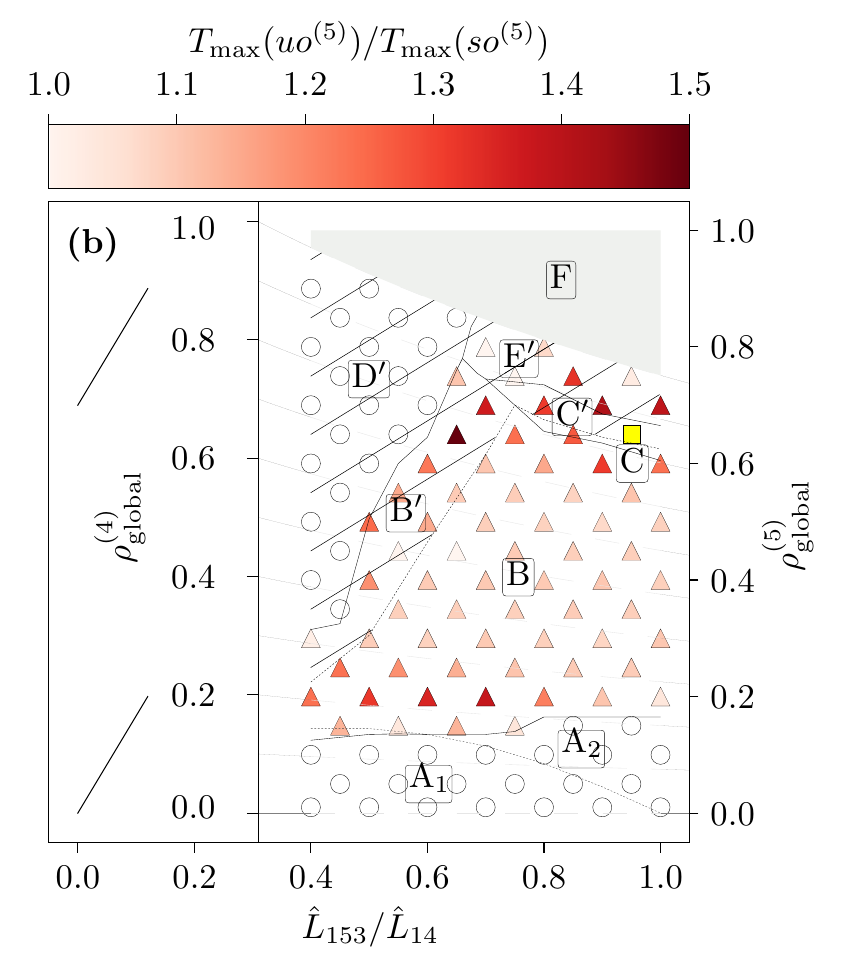}
  \includegraphics[width=0.49\columnwidth]{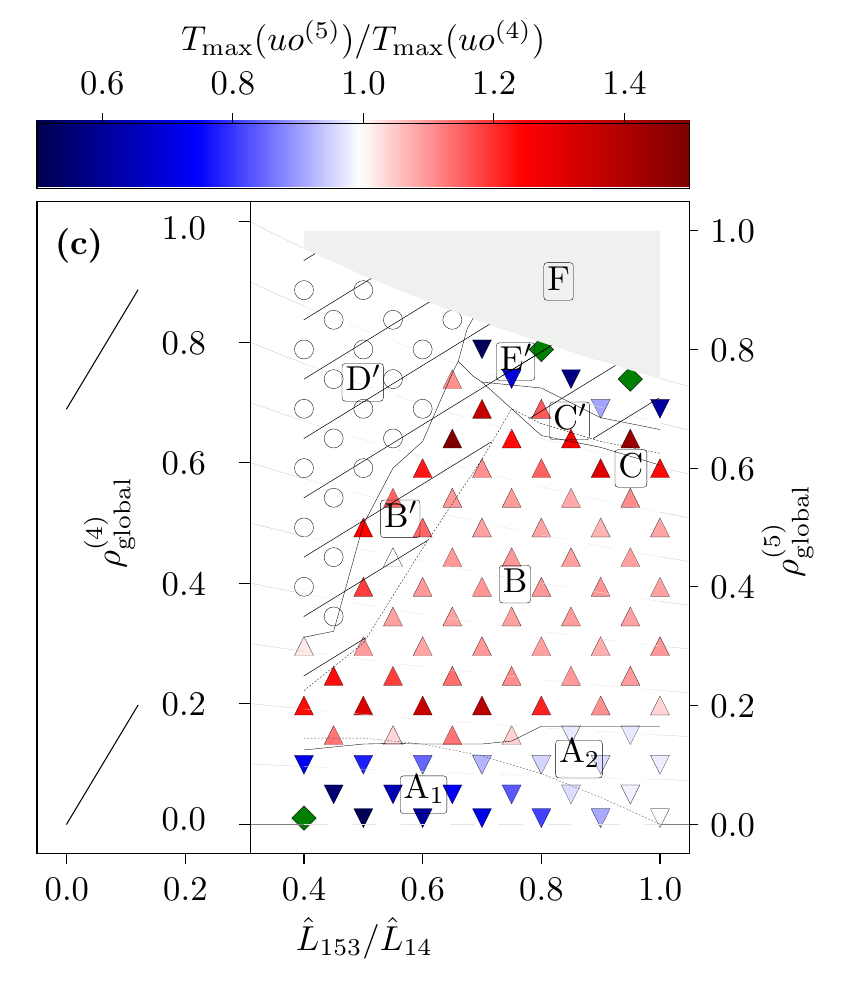}  
  \caption{\label{fig:closerlook}Quantifications of the influence of
    $E_5$ on the network. In all three parts, white $\bigcirc$
    indicate values equal to 1, coloured $\bigtriangleup$ indicate
    values between 1 and 1.5, coloured $\bigtriangledown$ indicate
    values between 0.5 and 1, green $\diamondsuit$ indicate values
    below 0.5 and yellow $\Box$ indicate values above 1.5. Part
    \textbf{(a)} shows how the new road could improve the system if it
    was always in its system optimum, measured by
    $T_{\text{max}}(so^{(5)})/T_{\text{max}}(so^{(5)})$. Part
    \textbf{(b)} measures the so-called price of anarchy given by
    $T_{\text{max}}(uo^{(5)})/T_{\text{max}}(so^{(5)})$. This measure
    explains how much the travel times go up in the 5link, if users
    are selfish and not guided by external measures. Part \textbf{(c)}
    shows how much the travel times go up/down due to $E_5$ compared
    to the 4link system if the network is used by selfish drivers,
    measured by $T_{\text{max}}(uo^{(5)})/T_{\text{max}}(uo^{(4)})$. }
\end{figure}

When talking about the Braess paradox, foremost one is interested in
the relation of the travel times of the user optima of a road network
before and after adding a new road. In Figures~\ref{fig:closerlook}
\textbf{(a)} to \textbf{(c)}, our two systems are studied in a bit
more detail by comparing $so^{(5)}$ to $so^{(4)}$, $so^{(5)}$ to
$uo^{(5)}$ and finally $uo^{(5)}$ to $uo^{(4)}$. Note that in parts
\textbf{(a)} and \textbf{(b)} of the figure, the values used for
$T_{\mathrm{max}}(so^{(5)})$ may actually not be the absolutely lowest
possible maximum travel times. This is due to the fact that, opposed
to the value of $\Delta T$ going to zero in the user optimum, a natural
lower bound for $T_{\mathrm{max}}$ does not exist. Therefore one
cannot be sure if the actual system optimum was found precisely 
(for more details, see~\ref{sec:app-sweep}).
Furthermore it is not guaranteed that we found all possible user
optima and therefore the values (given by the colors) should be
interpreted as tendencies of how the new road influences the system
and not as exact values (see~\ref{sec:app-special} for details - in
some cases actually multiple $uo$s were found).

In Figure~\ref{fig:closerlook} \textbf{(a)} additionally to the
underlying phase structure, for some specific points, the ratio of the
system optimum's maximum travel time in the 5link system and that of
the 4link system $T_{\text{max}}(so^{(5)})/T_{\text{max}}(so^{(4)})$
is shown.
In the 4link, the system optimum equals the user optimum and is thus
reached by selfish drivers without any traffic guidance. In the 5link
system, for selfish drivers the system optimum is only reached in the
$E_5$ optimal phase (phases A$_{1/2}$). In the other phases, the
system optimum would only be reached by forcing specific amounts of
particles to choose specific routes via some traffic guidance
system. In this case the system optimum could be reached
everywhere. Figure~\ref{fig:closerlook} \textbf{(a)} shows us these
potential benefits of route $E_5$. Since phase B (and B$^{\prime}$) is
a "Braess 1" phase and phase D$^{\prime}$ a "$E_5$ not used-like" phase,
here the system optima of 4link and 5link are equal and thus the ratio
is 1. In this large phase space region the new road cannot have any
positive effect - even if traffic is guided. The potential positive
effect of $E_5$ is largest for low global densities and low values of
$\hat{L}_{153}/\hat{L}_{14}$ (phase A$_1$) and for really high global
densities (phase E$^{\prime}$). In the first case, actually all cars
would use the new route and this would also be achieved without
traffic guidance since in this phase $uo^{(5)}=so^{(5)}$. The ratio
$T_{\text{max}}(so^{(5)})/T_{\text{max}}(so^{(4)})$ takes values as
low as 1/2 meaning that the $so$ travel time can be reduced by 1/2. In
phase E$^{\prime}$, the new route brings high benefits, since the
4link system here is almost full and thus travel times in the 4link
rapidly diverge (see Figure~\ref{fig:t4link}~\textbf{(b)}). Here the
increased capacity due to $E_5$ leads to high potential benefits,
$T_{\text{max}}(so^{(5)})/T_{\text{max}}(so^{(4)})$ takes values
smaller than 1/2. These full benefits would only be achieved by
traffic guidance in this case, since here $uo^{(5)}\neq so^{(5)}$.

In Figure~\ref{fig:closerlook} \textbf{(b)} the so-called price of
anarchy~\cite{pigou2013economics} inside the 5link system given by the
ratio of $T_{\text{max}}(uo^{(5)})/T_{\text{max}}(so^{(5)})$ is
shown. The price of anarchy is the ratio of the maximum travel time in
the user optimum (in phases A, B, C real $uo$ exist and thus all
travel times are almost equal in those $uo$) and the maximum travel
time in the system optimum
$T_{\text{max}}(uo^{(5)})/T_{\text{max}}(so^{(5)})$. In the 4link
system it is always equal to 1 since the user
optimum equals the system optimum. As already shown in
Figure~\ref{fig:possiblestates}, in the 5link system it is always
greater or equal to 1. In phases A$_{1/2}$, the '$E_5$ optimal'
phases, it is obviously equal to 1. In phase B, it is higher than 1
with values around 1.3. Not routing the traffic externally becomes
really costly at high densities in the 5link system. Here the maximum
travel times in the user optimum are in some cases more than 1.5 times
larger than those of the system optimum.

In Figure~\ref{fig:closerlook} \textbf{(c)} the ratio of the travel
times in the user optima of the 4link and 5link systems
$T_{\text{max}}(uo^{(5)})/T_{\text{max}}(uo^{(4)})$ is shown. This is
the situation that the original Braess paradox deals with:
how do the user optimum travel times in a system with selfish drivers
change due to the additional road. The new road is in the case of
selfish drivers only beneficial at low global densities and really
high global densities. In the "Braess 1" phase (phase B) it takes values
of approximately $1.2 - 1.4$ and the travel time is increased by
this factor due to $E_5$.


\section{Conclusion}

We analysed Braess' network with TASEP dynamics on the edges
considering periodic boundary conditions and individual drivers
following fixed individual strategies. Different to the case of
identical particles controlled by turning probabilities at
the junction sites~\citep{PhysRevE.94.062312} we could find stable
travel times throughout the whole density region of both the network
with and without the new edge. Without the new road, system and user 
optima could be found for all densities. In the major
part of the phase space real user optima could also be identified in
the system with the new road. Only if the new route resulting from the
new road is really short compared to the old routes or the global
density is really high, no real user optima exist.

In the phase space region where real user optima exist, the new road
leads to lower travel times only at low global densities
$0 \leq \rho_{\mathrm{global}}^{(4)/(5)} \lesssim 0.2$. The largest
part of the phase space is dominated by the Braess phases, with the
"Braess 1" phase being the most prominent phase. The "Braess 2" phase
is also found in a small phase space region.

At high global densities or small path lengths of the new path no real
user optima exist. This is due to the fact that here the system is
prone to gridlock completely if all particles try to
reduce their travel times. In this region of the
phase space we chose the closest candidate of a user optimum without a
possibility of gridlocks to construct the phase diagram. It turns out
that here the new road also leads to higher travel times in most
situations or is even completely ignored. Only at really high
densities the road leads to lower travel times. Still in this whole
region the system is at risk of gridlocks. This gridlock risk is also
a consequence of added periodic boundary conditions and could vanish
if the system was treated with open boundaries and a sufficiently large
exit probability.

Additionally to constructing the phase diagram we could quantify how
the new road influences the system for selfish users and also what
would happen if an external travel guidance authority would drive the
system into its system optimum. Even if traffic was guided the
network's performance, in terms of tavel times, would only be improved
at really high or really low densities. The price of anarchy is
relatively low inside the system with the new road.

In the present system the negative implications of the new road on the network
performance are even more dominant than in the version
of the model at a higher level of stochasticity which we studied
before~\cite{PhysRevE.94.062312}. It provides further evidence for the
claim that Braess' paradox is of major importance when studying
networks of microscopical traffic models. This hints at the paradox's
importance in real road networks. When building new connections in a
road network with the aim of reducing travel times, city planners
should be very careful. Also one can see that the route choice
behaviour can have enormous effects on the traffic situation. Even if
on average both mechanisms in the present paper and
in~\citep{PhysRevE.94.062312} are very similar, the fixed strategies
lead to a different behaviour of the system and the disappearance of
one of the most dominant phases observed in the system with turning
probabilities.

\section*{Acknowledgements}
Financial support by Deutsche Forschungsgemeinschaft (DFG) under grant
SCHA~636/8-2 is gratefully acknowledged. Also support of conference
travel expanses by the Bonn-Cologne Graduate School of Physics and
Astronomy (BCGS) is acknowledged.  Monte Carlo simulations were
carried out on the CHEOPS (Cologne High Efficiency Operating Plattform
for Science) cluster of the RRZK (University of Cologne).

\clearpage
\appendix

\section{Measurement process} 
\label{sec:app-measurement-process} 

Here we describe how travel times in both the 4link and 5link systems
were measured and how we proceeded to find user and system optima. In
all our measurements, lengths were chosen as $L_0=1$, $L_1=L_3=100$,
$L_2=L_4=500$ and $L_5$ and the total number of particles $M$ were
varied. In all Monte Carlo simulations, random numbers were generated
using the Mersenne Twister algorithm.  The system was initialized
randomly but particles are placed according to their strategies,
i.e. randomly on the routes they want to use. A particle of strategy
153 could for example not be placed on edge $E_2$.  This kind of
initialization was chosen over a completely random initialization to
avoid gridlocks due to wrong initialization. After initialization, the
system was always relaxed for at least $5\times 10^5$ sweeps. Each
particle has its own fixed strategy and for each particle and all of
its finished rounds, the travel times were measured.  Note that this
is a different measurement process than in our previous
article~\citep{PhysRevE.94.062312}. To obtain travel time values the
system was kept running until the desired amount of measurements for
each route was gathered.

\subsection{Number of measurements}

\begin{figure}[h!]
  \centering
  \includegraphics[width=0.49\columnwidth]{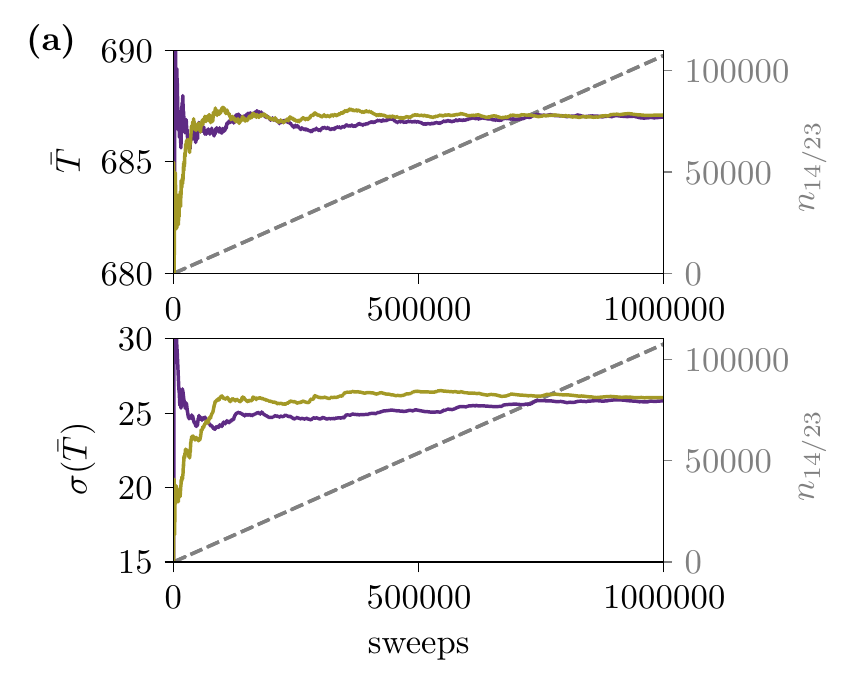}   
  \includegraphics[width=0.49\columnwidth]{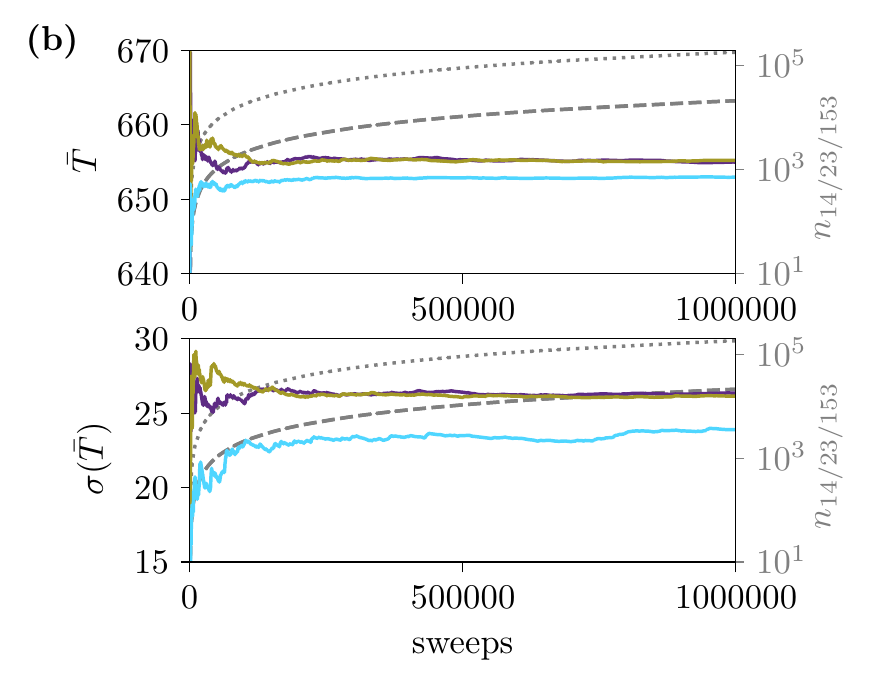}
  \caption{\label{fig:tentwicklung} The development of the mean
    $\bar{T}$ and the standard deviation $\sigma(\bar{T})$ of the
    travel time values against the number of sweeps. The values for
    routes 14/23/153 are shown in
    \textcolor{purple-density}{purple}/\textcolor{brown-density}{brown}/\textcolor{blue-density}{blue}. Part
    \textbf{(a)} shows measurements of the 4link system for $M=148$
    and $n_{\text{l}}^{(j_1)}=0.5$ and part \textbf{(b)} shows
    measurements of the 5link system with $L_5=278$ and $M=148$ and
    $n_{\text{l}}^{(j_1)}\approx 0.92$ and
    $n_{\text{l}}^{(j_2)}\approx 0.10$. The second y-axes show how
    many measurements $n_{14/23/153}$ of $T_{14/23/153}$ were
    performed. The curve with the long dashes represent the number of
    measurements of the travel times of route 14 and 23, the curve
    with the smaller dashes the number of measurements of the travel
    time of route 153.}
\end{figure}

To obtain travel time values for the routes, the travel times have to
be measured sufficiently often. In Figure~\ref{fig:tentwicklung} it is
shown how the mean value\footnote{Note that in the main part of this
  article, when talking about measured travel time values, those are
  always the mean values of many measurements!} and standard deviation
(Equation~(\ref{eq:sd}) multiplied by $\bar{T}$) of the travel times
develop with the number of measurements for an example parameter set.

\clearpage

\subsection{Sweeping the network state - landscapes to find user and
  system optima} 
\label{sec:app-sweep}
\begin{figure}[h!]
  \centering
  \includegraphics[width=0.49\columnwidth]{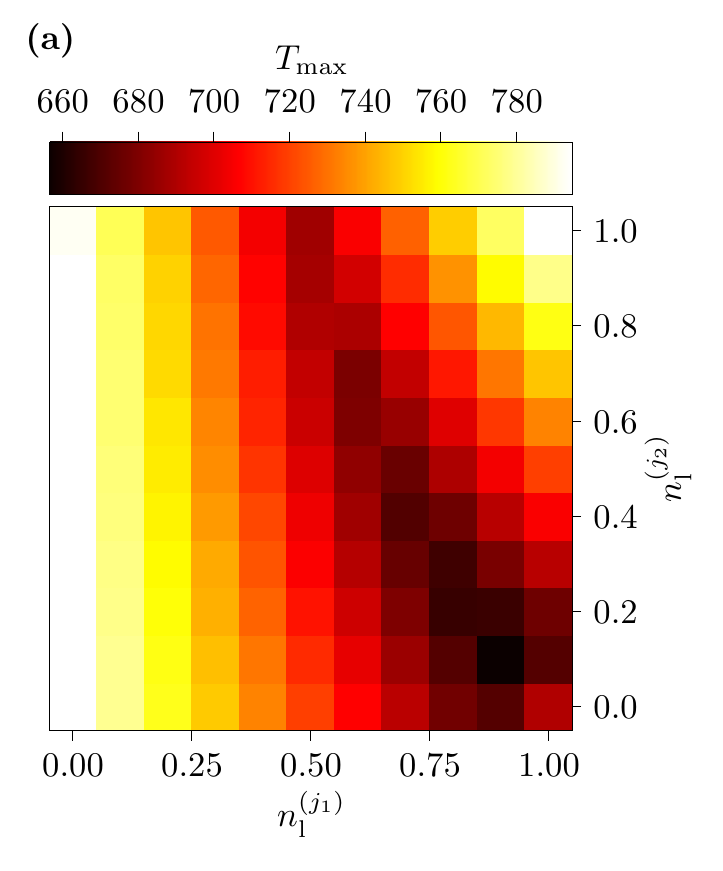}
  \includegraphics[width=0.49\columnwidth]{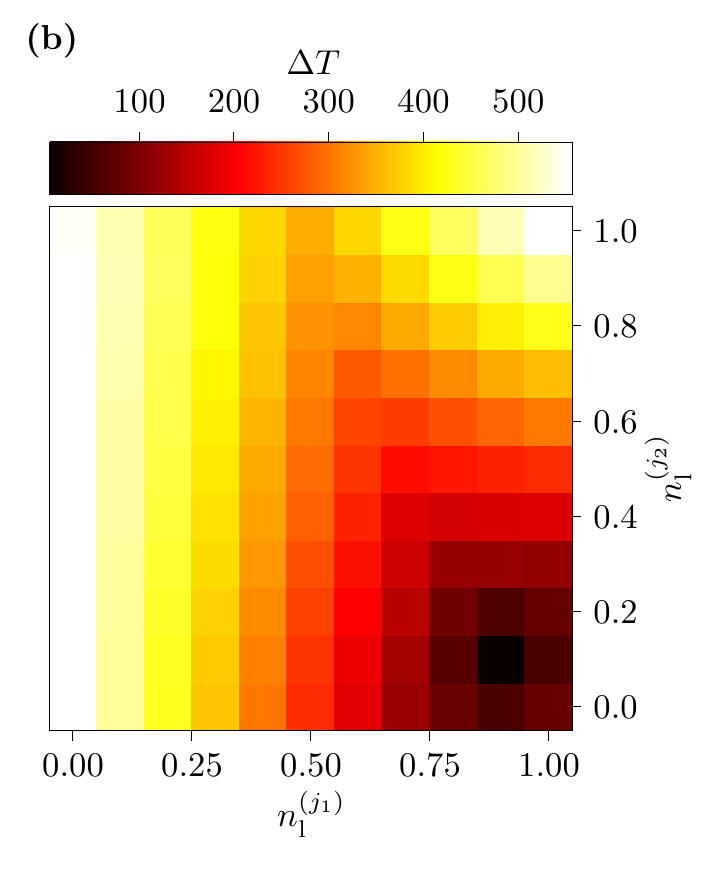}
  \caption{\label{fig:sweep-landscape}The $\Delta T$ and
    $T_{\text{max}}$ landscapes of the 5link system for $L_5=278$ and
    $M=148$ depending of the $n_{\text{l}}^{(j_1)}$ and
    $n_{\text{l}}^{(j_2)}$. As can be seen, this is an "$E_5$ optimal"
    state, since the minima of $\Delta T$ and $T_{\text{max}}$
    coincide at $n_{\text{l}}^{(j_1)}=0.9$ and $n_{\text{l}}^{(j_2)}=0.1$
    and the corresponding strategy has a lower maximum travel time
    than the 4link's system optimum which is found at
    $n_{\text{l}}^{(j_1)}=0.5$ and $n_{\text{l}}^{(j_2)}=1.0$.}
\end{figure}

A simple yet CPU time costly way to find the user and system optima is
to check all different combinations of $N_{14}$, $N_{23}$ and
$N_{153}$, or all combinations $n_{\text{l}}^{(j_1)}$ and
$n_{\text{l}}^{(j_2)}$, both varying from 0 to 1, for a given
$(L_5, M)$ and see, which combinations minimize $\Delta T$ and
$T_{\text{max}}$ respectively. We used a grid resolution of 0.1 for
finding the system optima of the 5link system. An example of how the
$\Delta T$ and $T_{\text{max}}$ landscapes look like can be seen in
Figure~\ref{fig:sweep-landscape}. It should be noted that the actual
system optima may in most cases lie in between the 0.1 grid
points. Nevertheless, this grid is sufficient to see whether the
system optimum of the 5link system has a lower or higher maximum
travel time than that of the 4link system for a given $(L_5, M)$. It
has to be noted though that this technique does not ensure that we
found the actual system optima, which is why
the coloured points in Figures~\ref{fig:closerlook} \textbf{(a)} and
\textbf{(b)} should not be seen as accurate values. For finding user
optima we used a more accurate method as described in~\ref{sec:MC}.

\subsection{Metropolis algorithm for finding user optima} 
\label{sec:MC}

\begin{figure}[h!]
  \centering
  \includegraphics[width=0.49\columnwidth]{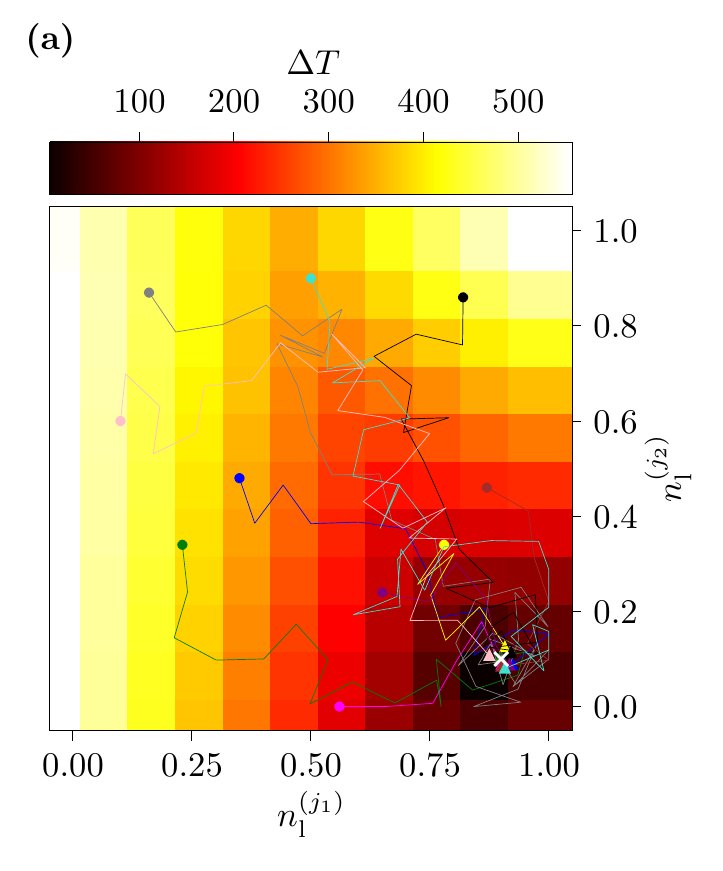}
  \includegraphics[width=0.49\columnwidth]{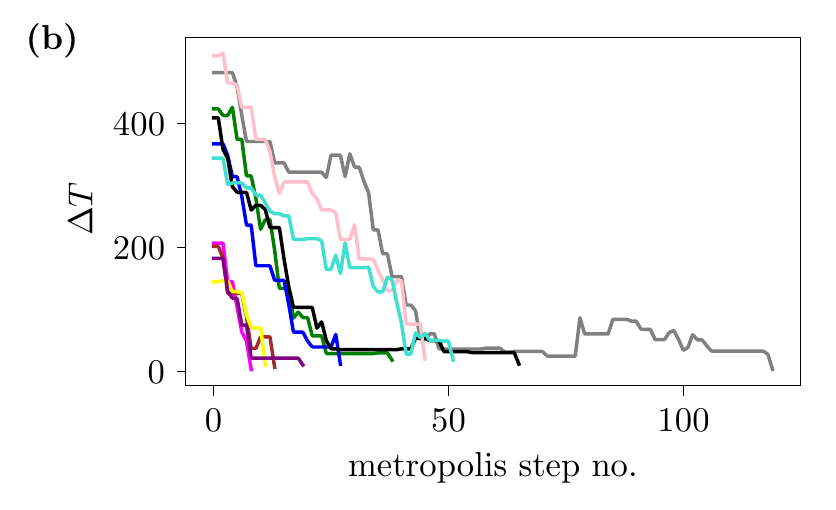}
  \caption{\label{fig:mc-verlauf}An example of the performance of the
    Metropolis algorithm for $L_5=278$ and $M=148$ and ten different
    start values. In \textbf{(a)} the search paths are shown with
    underlayed values of $\Delta T$ which were obtained by sweeping
    the $(n_{\text{l}}^{(j_1)}, n_{\text{l}}^{(j_2)})$-landscape as
    described in~\ref{sec:app-sweep}. The beginnings of all paths are
    marked by a $\bigcirc$ and the endings by a $\bigtriangleup$. Also
    the user optimum is marked by a white $\times$. In \textbf{(b)}
    the corresponding $\Delta T$ values against the number of
    Metropolis steps are shown. One can see, that the algorithm
    converges pretty fast for all 10 start values.}
\end{figure}

The method of sweeping the $\Delta T$ and $T_{\text{max}}$ landscapes
to find user optima, as described in~\ref{sec:app-sweep}, is not very
effective. If the whole region $n_{\text{l}}^{(j_1)}$ and
$n_{\text{l}}^{(j_2)}$ both from 0 to 1 is sweeped, a lot of
measurements are performed far away from the user (and system)
optima. Those measurements are a waste of CPU time. Furthermore,
depending on how fine the step size is, the user optima will most
likely not lie on one of the scanned grid points but in between. A
brute force way would be rescanning the corresponding landscape
regions with a finer grid. Since this is very time intensive we
developed a Metropolis Monte-Carlo~\cite{doi:10.1063/1.1699114}
algorithm for finding user optima. The algorithm works like this:
\begin{enumerate}
 \item Set maximum step width $sw$ and 'temperature' $T$
 \item Set start values $(n_{\text{l}}^{(j_1)}, n_{\text{l}}^{(j_2)})$ and from this 
 \begin{itemize}
   \item $N_{14}=M\cdot n_{\text{l}}^{(j_1)}\cdot n_{\text{l}}^{(j_2)}$ 
   \item $N_{23}=M\cdot(1-n_{\text{l}}^{(j_1)})$
   \item $N_{153}=M\cdot n_{\text{l}}^{(j_1)}\cdot (1-n_{\text{l}}^{(j_2)})$
  \end{itemize}
\item Let the system thermalize with strategy according to
  $(n_{\text{l}}^{(j_1)}, n_{\text{l}}^{(j_2)})$
 \item Measure travel times $T_{14}$, $T_{23}$, $T_{153}$ and calculate $\Delta T$
 \item Suggest new
   $(n_{\text{l, new}}^{(j_1)}, n_{\text{l,new}}^{(j_2)})$ by drawing
   a random number $z$ between 0 and 2$\pi$ and setting
   $(n_{\text{l, new}}^{(j_1)}, n_{\text{l,
       new}}^{(j_2)})=(n_{\text{l}}^{(j_1)}+sw\cdot \cos(z),
   n_{\text{l}}^{(j_2)}+sw\cdot \sin(z))$
   and calculate $N^{\text{new}}_{14}$, $N^{\text{new}}_{23}$,
   $N^{\text{new}}_{153}$ as in step 2
  \item Let the system thermalize with strategy according to $(n_{\text{l, new}}^{(j_1)}, n_{\text{l, new}}^{(j_2)})$
  \item Measure travel times $T^{\text{new}}_{14}$, $T^{\text{new}}_{23}$, $T^{\text{new}}_{153}$ and calculate $\Delta T^{\text{new}}$
  \item Accept the new strategy with probability $p=\min\left( 1, \exp\left(-\frac{\Delta T-\Delta T^{\text{new}}}{T}\right)\right)$
  \item Repeat steps 5 to 7 as long as
    $\Delta T^{\text{new}}>\epsilon$, with tolerance $\epsilon$
\end{enumerate}
In this algorithm, the maximum step width $sw$ is the maximum possible
value, $n_{\text{l}}^{(j_1)}$ and $n_{\text{l}}^{(j_2)}$ can be
changed by. The temperature $T$ is a measure for the probability with
which a strategy with higher $\Delta T$ might be accepted and
$\epsilon$ is the tolerance: if $\Delta T \leq \epsilon$ the strategy
is accepted as the user optimum. The 'real' user optimum is reached,
if $\epsilon$ is exactly zero. It turned out to be useful to set
$\epsilon=20$, $sw=0.1$ and $T=10$. An additional tenth step could be
added to the algorithm, in which $sw$ would be reduced, if newly
suggested probabilities get rejected a certain amount of times. This
additional step was not needed in our case. The algorithm performed
well in most of our situations. Fig.~\ref{fig:mc-verlauf}~\textbf{(a)}
shows the search path of the algorithm for $L_5=278$ and $M=148$ for
10 different start values
$(n_{\text{l}}^{(j_1)}, n_{\text{l}}^{(j_2)})$.  The $uo$-landscape
with 0.1 step width as described in the previous subsection is
underlayed for visualization purposes.
Furthermore, in Fig.~\ref{fig:mc-verlauf}~\textbf{(b)} the $\Delta T$
values against the Metropolis step nummber (i.e. how often steps 5 to
8 of the algorithm were performed) is shown. From both pictures it can be
deduced that the algorithm works really well for this case. Indeed,
it performed really well for most other $(L_5, M)$ combinations as
well.  Sometimes, depending on the start values, the algorithm will
not converge and has to be restarted with different start values. Also
in some cases there is more than one user optimum, as described
in~\ref{sec:app-special}. In this case, the algorithm will get
'caught' in one of them. The algorithm can also be used to find system
optima if after each step $T_{\text{max}}$ is calculated and the
newly suggested strategy is accepted if $T_{\text{max}}$ got
lower. The problem in this case is that there is no real termination
condition as there is no a priori known lower bound to
$T_{\text{max}}$.


\section{Special cases} 
\label{sec:app-special}

\begin{figure}[h]
  \centering
  \includegraphics[width=0.49\columnwidth]{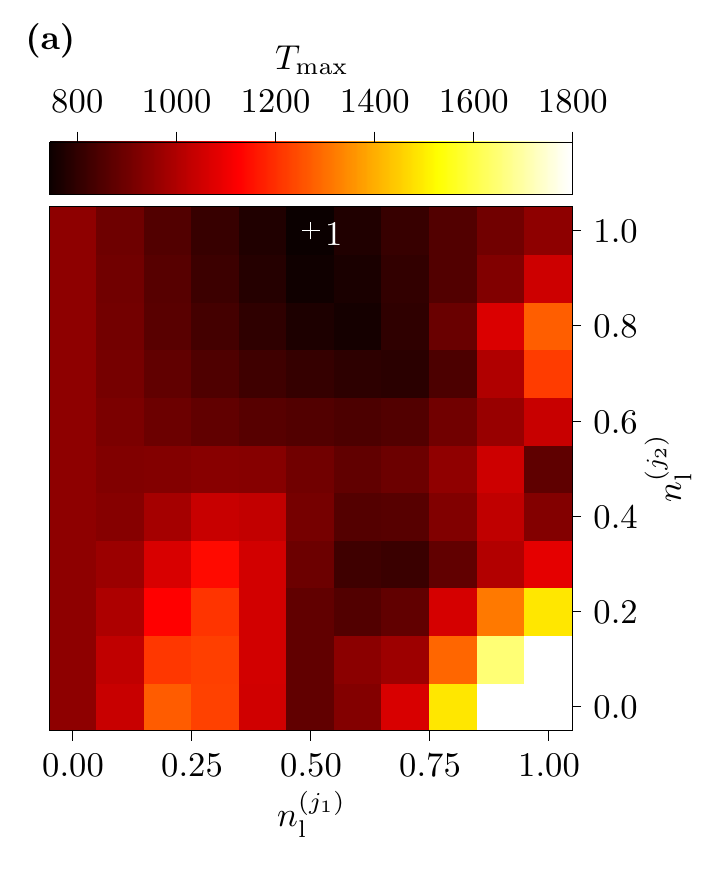}
  \includegraphics[width=0.49\columnwidth]{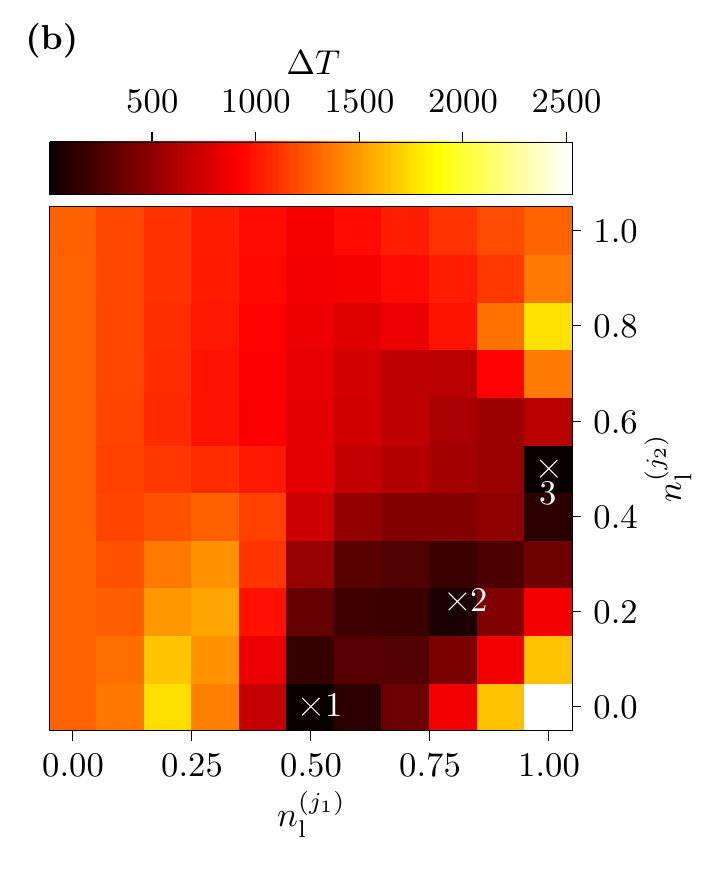}
  \caption{\label{fig:specialcases}The $T_{\mathrm{max}}$ (part
    \textbf{(a)}) and $\Delta T$ (part \textbf{(b)}) landscapes of the
    5link system with $L_1=L_3=100$, $L_2=L_4=500$, $L_5=37$, $M=224$,
    $\rho^{(5)}_{\text{global}}=0.18$. In part \textbf{(a)}, the
    system optimum is marked by $+ 1$. In part \textbf{(b)} one can
    see that there are three different user optima, $\times 1$ to
    $\times 3$, the values of which are given in
    Table~\ref{tab:special}.}
\end{figure}

For some configurations we could find more than one user optimum. Here
we present the example of $\hat{L}_{153}/\hat{L}_{14}=0.4$ and
$\rho^{(5)}_{\text{global}}=0.18$. In Figure~\ref{fig:specialcases}
the $T_{\mathrm{max}}$ and $\Delta T$ landscapes for this parameter
set are shown. In the figure, the values from a sweep of the
landscapes in steps of 0.1 is underlayed.
The travel time and $\Delta T$ and $T_{\mathrm{max}}$ values for the
four marked points are given in Table~\ref{tab:special}. From part
\textbf{(a)} of the picture one can see that the system optimum is
given by $n_{\text{l}}^{(j_1)}=0.5$ and
$n_{\text{l}}^{(j_2)}=1.0$. This means that here $so^{(5)}=so^{(4)}$
since this is the state where half of the particles use route 14 and
the other half route 23.

In part \textbf{(b)} of the Figure one can see that there are three
different user optima. The two user optima $\times 1$ and $\times 3$
were found by sweeping the $n_{\text{l}}^{j_1/j_2}$ and are special
cases of user optima which were already mentioned in
Section~\ref{sec:mc-observables}. The user optimum $\times 2$ was
found by our Metropolis algorithm. The optimum $\times 1$ is a special
case since only paths 23 and 153 are used.  Since both their travel
times are almost equal and smaller than that of the unused route 14
this state is a user optimum. For calculating $\Delta T$, only the
difference between $T_{23}$ and $T_{153}$ is used. It would in this
case not make sense for any particle to switch to route 14 which has a
higher travel time. The same happens in the user optimum $\times 3$,
but here routes 14 and 153 are used and route 23 is not. The other
user optimum $\times 2$ is an 'ordinary' user optimum in which all
three routes are used and have (almost) the same travel time.
\begin{table}[h]
  \caption{\label{tab:special} The $n_{\text{l}}^{(j_1/j_2)}$, 
    $\Delta T$, $T_{\mathrm{max}}$ and travel time values of the routes for the 4 points which are marked in Figure~\ref{fig:specialcases}. The points $\times 1$ to $\times 3$ are three user optima of the system, while $+ 1$ is the system optimum.}
\begin{center}
    \begin{tabular}{| c | l | l | l | l | l | l | l |}
     \hline
    Point &  $n_{\text{l}}^{(j_1)}$ & $n_{\text{l}}^{(j_2)}$ & $T_{14}$ & $T_{23}$ & $T_{153}$ &  $\Delta T$ & $T_{\mathrm{max}}$ \\ \hline
    $\times 1$ & 0.5 &  0.0 & 926 & 880 & 876 & 4 & 880 \\ \hline
    $\times 2$ & 0.808 & 0.221 &  970 & 975 & 975 & 10 & 975 \\ \hline
    $\times 3$ & 1.0 & 0.5 & 878 & 1136 & 875 & 3 & 878  \\ \hline\hline
    $+ 1$ & 0.5 &  1.0 & 743 & 742 & 294 & 898 & 743 \\ \hline
    \end{tabular}
\end{center}
\end{table}
The (maximum) travel time in all three user optima is higher than that
of the system optimum (which is the same as the 4link system optimum)
which leads to the conclusion that no matter in which user optimum
the system ends up, a "Braess 1" state is present.

While in the whole unhatched area of the phase diagram
(Figure~\ref{fig:phases}) user optima were found we cannot guarantee
that \textit{all} user optima were found. Also, the values (given by
the color of the points) in Figures~\ref{fig:closerlook} \textbf{(a)}
to \textbf{(c)} are given for one of the found user optima and could
be slightly different if, for the cases of multiple user optima,
another user optimum was chosen. This is because the user optima, as
in the example presented here (see Table~\ref{tab:special}), could
have different travel time values. Therefore the values in
Figure~\ref{fig:closerlook} should not be interpreted as exact values
but as a tendency of how the system changes due to the addition of
$E_5$.

The fact that we found multiple user optima with different travel
times (and different $T_{\mathrm{max}}$ values and also different
total travel time values) for the same parameter set is also a
difference to what is observed in mathematical models of road traffic. In
these models it was shown that "$[$the user optimum$]$ is unique
whenever the shortest routes between all pairs of locations are unique
and cost is strictly increasing with increasing
flow"~\cite{beckmann1956studies}.

\clearpage

\section*{References}

\bibliography{paper-braess-2}

\end{document}